\documentclass[superscriptaddress, 10pt, prx, aps, twocolumn,longbibliography]{revtex4-1}
\usepackage{graphicx}

\usepackage{bm}
\usepackage{bbm}

\usepackage{amsmath}
\usepackage{amsthm}
\usepackage{amssymb}

\usepackage{tabularx}
\usepackage{gensymb}
\usepackage{color}
\usepackage[dvipsnames]{xcolor}
\usepackage[normalem]{ulem}
\usepackage{xspace}

\newtheorem{definition}{Definition}
\newtheorem{result}{Result}
\newtheorem{theorem}{Theorem}
\newtheorem{lemma}{Lemma}

\makeatletter
\DeclareRobustCommand{\cev}[1]{%
  \mathpalette\do@cev{#1}%
}
\newcommand{\do@cev}[2]{%
  \fix@cev{#1}{+}%
  \reflectbox{$\m@th#1\vec{\reflectbox{$\fix@cev{#1}{-}\m@th#1#2\fix@cev{#1}{+}$}}$}%
  \fix@cev{#1}{-}%
}
\newcommand{\fix@cev}[2]{%
  \ifx#1\displaystyle
    \mkern#23mu
  \else
    \ifx#1\textstyle
      \mkern#23mu
    \else
      \ifx#1\scriptstyle
        \mkern#22mu
      \else
        \mkern#22mu
      \fi
    \fi
  \fi
}

\DeclareMathAlphabet\mathbfcal{OMS}{cmsy}{b}{n}
\newcommand{\past}[1]{\cev{#1}}
\newcommand{\future}[1]{\vec{#1}}
\newcommand{\omni}[1]{\overleftrightarrow{#1}}

\newcommand{\ph}{P_h}
\newcommand{\Ph}{\mathcal{P}_h}
\usepackage{enumitem}
\allowdisplaybreaks[1]

\newcommand{\id}[1]{\mathbb{1}}

\DeclareMathOperator{\tr}{Tr}

\newcommand{\Lx}{\past{X}}
\newcommand{\Rx}{\future{X}}

\newcommand{\bra}[1]{\langle #1|}
\newcommand{\ket}[1]{|#1\rangle}
\newcommand{\braket}[2]{\langle #1|#2\rangle}
\newcommand{\ketbra}[2]{|#1\rangle\!\langle#2|}

\begin{document}
\title{Causal Asymmetry in a Quantum World}
\author{Jayne Thompson}
\email{thompson.jayne2@gmail.com}
\affiliation{Centre~for~Quantum~Technologies, National~University~of~Singapore, 3 Science Drive 2, 117543, Singapore}

\author{Andrew~J.~P.~Garner}
\affiliation{Centre~for~Quantum~Technologies, National~University~of~Singapore, 3 Science Drive 2, 117543, Singapore}
\affiliation{Institute for Quantum Optics and Quantum Information, Austrian Academy of Sciences, Boltzmanngasse 3, Vienna, A-1090, Austria }

\author{John R. Mahoney}
\affiliation{Complexity Sciences Center and Physics Department,
University of California at Davis, One Shields Avenue, Davis, CA 956}

\author{James P. Crutchfield}
\affiliation{Complexity Sciences Center and Physics Department,
University of California at Davis, One Shields Avenue, Davis, CA 956}

\author{Vlatko Vedral}
\affiliation{Atomic~and~Laser~Physics, University~of~Oxford, Clarendon~Laboratory, Oxford, OX1 3PU, United Kingdom}
\affiliation{Centre~for~Quantum~Technologies, National~University~of~Singapore, 3 Science Drive 2, 117543, Singapore}
\affiliation{Department of Physics, National University of Singapore, 3 Science Drive 2, Singapore 117543}

\author{Mile Gu}
\email{gumile@ntu.edu.sg}
\affiliation{School of Physical and Mathematical Sciences, Nanyang Technological University, Singapore 639673,  Singapore}
\affiliation{Complexity Institute, Nanyang Technological University, Singapore 639673,  Singapore}
\affiliation{Centre~for~Quantum~Technologies, National~University~of~Singapore, 3 Science Drive 2, 117543, Singapore}

\begin{abstract}

Causal asymmetry is one of the great surprises in predictive modelling: the memory required to predict the future differs from the memory required to retrodict the past. There is a privileged temporal direction for modelling a stochastic process where memory costs are minimal. Models operating in the other direction incur an unavoidable memory overhead. Here we show that this overhead can vanish when quantum models are allowed. Quantum models forced to run in the less natural temporal direction not only surpass their optimal classical counterparts, but also any classical model running in reverse time. This holds even when the memory overhead is unbounded, resulting in quantum models with unbounded memory advantage.
\end{abstract}

\maketitle

\begin{figure}
\includegraphics[width=0.9\columnwidth]{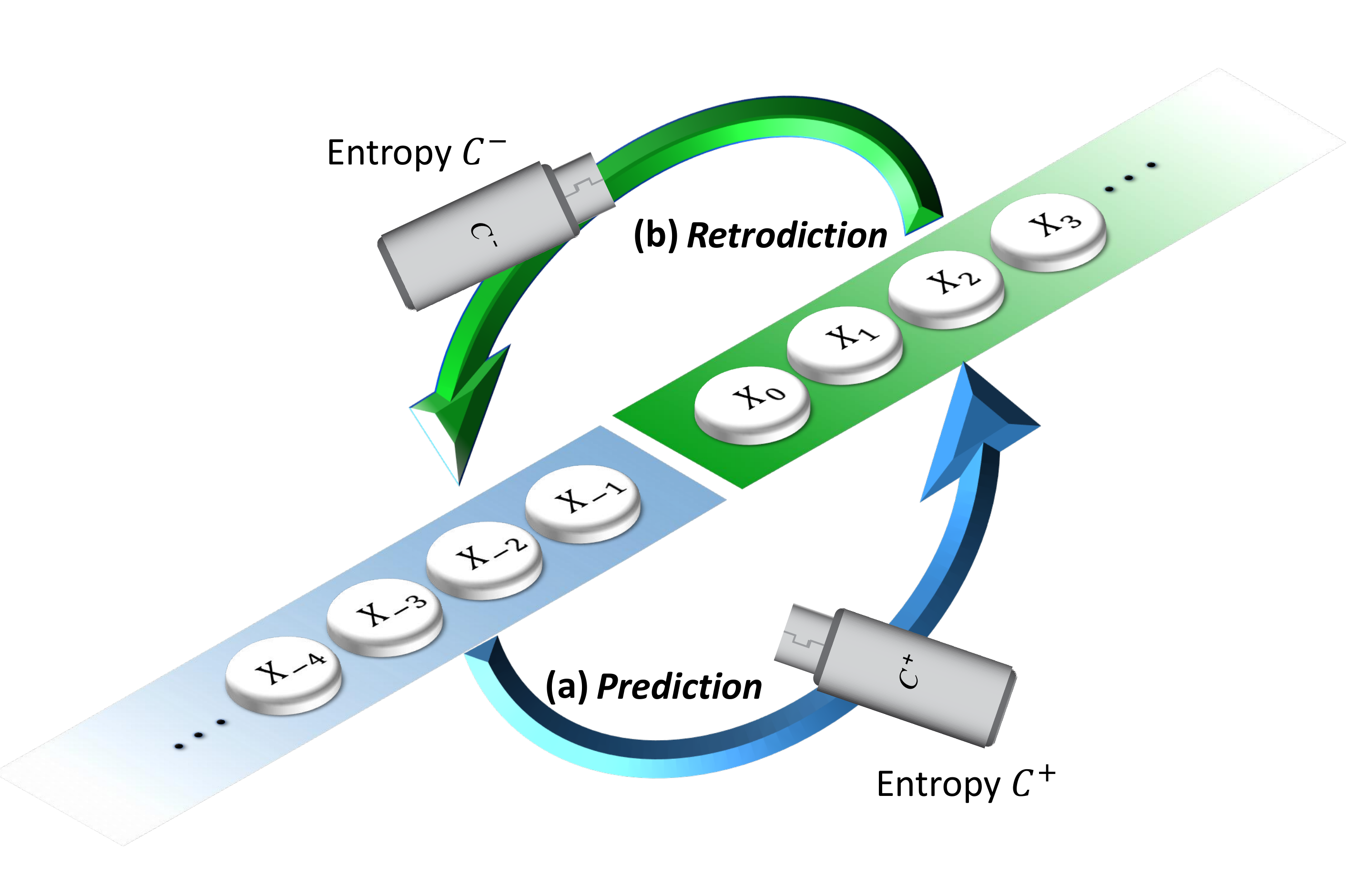}\caption{A stochastic process can be modeled in either temporal order. (a) A causal model takes information available in the past $\past{x}$ and and uses it to make statistically accurate predictions about the process' conditional
future behaviour $P(\future{X} | \past{X} = \past{x})$. (b) A retrocausal model replicates the system's behaviour, as seen by an observer who scans the outputs from right to left encountering $X_{t+1}$ \emph{before} $X_{t}$. Thus it stores relevant future information $\future{x}$, in order to generate a statistically accurate retrodiction of the past $P(\past{X} | \future{X} = \future{x})$.  Causal asymmetry implies a non-zero gap between the minimum memory required by any causal model $C^+$, and its retrocausal counterpart $C^-$. \label{fig:timereversal}}
\end{figure}

How can we observe an asymmetry in the temporal order of events when physics at the quantum level is time-symmetric? The source of time's barbed arrow is a longstanding puzzle in foundational science \cite{linden2009quantum,popescu2006entanglement,page1983evolution,price1997time}. Causal asymmetry offers a provocative perspective \cite{crutchfield2009time}. It asks how Occam's razor -- \emph{the principle of assuming no more causes of natural things than are both true and sufficient to explain their appearances} -- can privilege one particular temporal direction over another. That is, if we want to model a process causally -- such that the model makes statistically correct future predictions based only on information from the past --  what is the minimum past information we must store? Are we forced to store more data if we model events in one particular temporal order over the other (see Fig. \ref{fig:timereversal})?

Consider a cannonball in free fall. To model its future trajectory, we need only its current position and velocity. This remains true even when we view the process in reverse-time. This exemplifies causal symmetry. There is no difference in the amount of information we must track for prediction versus retrodiction. However this is not as obvious for more complex processes. Take a glass shattering upon impacting the floor. In one temporal direction, the future distribution of shards depends only on the glass's current position, velocity and orientation. In
the opposite, we may need to track relevant information regarding each glass shard to infer the glass's
prior trajectory. Does this require more or less information? This potential divergence is quantified in the theory of computational mechanics~\cite{crutchfield2012between}. It is not only generally non-zero, but can also be unbounded. This phenomenon implies a simulator operating in the `less natural' temporal direction is penalized with potentially unbounded memory overhead, and is cited as a candidate source of time's barbed arrow~\cite{crutchfield2009time}.

These studies assumed that all models are implemented using classical physics. Could the observed causal asymmetry have been a consequence of this classicality constraint? Here, we first consider a particular stochastic process that is causally asymmetric. We determine the minimal information needed to model the same process in forward versus reverse time using quantum physics, and prove these quantities exactly coincide. More generally, we present systematic methods to model any causally asymmetric stochastic process quantum mechanically. Critically, the resulting quantum models not only use less information than any classical counterpart, but also any classical model of the time-reversed process. Thus, quantum models can field a memory advantage, that always exceeds the memory overhead incurred by causal asymmetry. Our work indicates this overhead can emerge when imposing classical causal explanations.
These result remain true even in cases where causal asymmetry becomes unbounded.

\section{Background}

\textbf{Framework} -- Consider a system that emits an output $x_t$ governed by some random variable $X_t$ at each discrete point in time $t$. This behaviour can be described by a stochastic process $\mathcal{P}$ -- a joint probability distribution $P(\past{X},\future{X})$ that correlates past behaviour, $\past{X} = \dots X_{-2} X_{-1}$, with future expectations, $\future{X} = X_{0} X_{1} \dots$. Each instance of the past $\past{x} = \dots x_{-2} x_{-1}$ exhibits a conditional future  $\future{x} = x_{0} x_{1}\dots$ with probability $P(\future{X} = \future{x} | \past{X} = \past{x})$.

Suppose that a model for this system can replicate this future statistical behaviour using only $H$ bits of past information. Then this model can be executed by encoding the past $\past{x}$ into a state $s(\past{x}) \in \mathcal{S}$ of a physical system $\Xi$ of entropy $H$, such that repeated application of a systematic action  $\mathcal{M}$ on $\Xi$  sequentially generates  $x_0$,$x_1 \ldots$ governed by the conditional future $P(\future{X}|\past{X} = \past{x})$.
The model is \emph{causal} if at each instance of time, all the information $\Xi$ contains about the future can be obtained from the past~\footnote{Formally, this implies that the conditional mutual information $I(R,\future{X}|\past{X})$ is zero, where $R$ is the random variable governing the state of $\Xi$. In literature, $I(R,\future{X}|\past{X})$ is named oracular information, as it captures the amount of information a state of the memory $s$ stores about the future that is not contained in the past~\cite{ellison2011information, crutchfield2010synchronization}. Note that this notion of causal model differs from the recent developments in quantum causal models \cite{allen2017quantum,costa2016quantum} and foundation work on causal inference \cite{pearl2009causal}.}. Implementing it on a computer then gives us a statistically faithful simulation of the process'
realizations. The simplest causal model for a process $P(\past{X}, \future{X})$, is the model that minimizes $H$.

The \emph{statistical complexity} $C^+$ is defined as the entropy $H$ of this simplest model -- it is the minimal amount of past information needed to make statistically correct future predictions~\cite{crutchfield1989inferring,shalizi2001computational}. This measure is used to quantify structure in diverse settings~\cite{Gonçalves1998385,park2007complexity,tino1999extracting}, including hidden variable models emulating quantum contextuality~\cite{cabello2018optimal}. $C^+$ also fields thermodynamic significance, having been linked to the minimal heat dissipation in stochastic simulation and the minimal structure a device needs to fully extract
free energy from non-equilibrium environments~\cite{garner2017thermodynamics,wiesner2012information,cabello2016thermodynamical,boyd2017transient}.

\begin{figure*}
\includegraphics[width=0.8\textwidth]{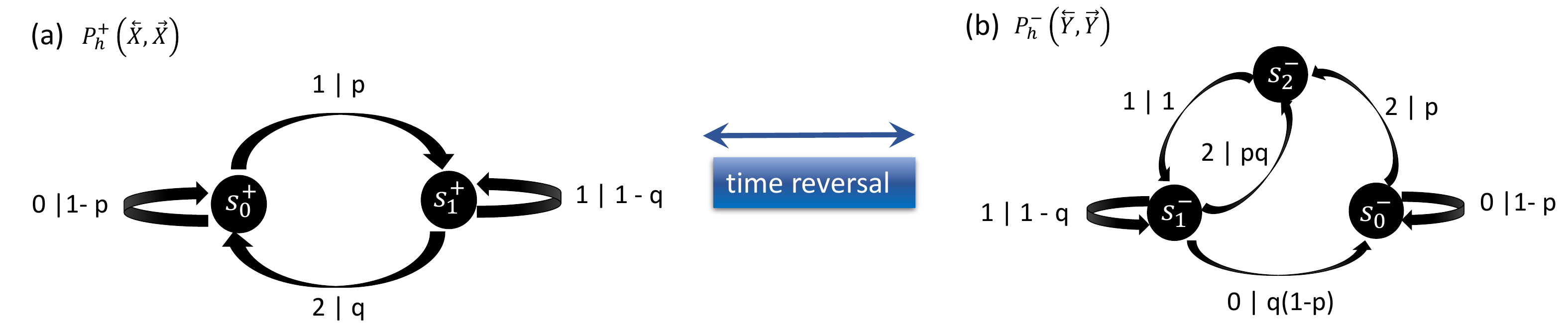}\caption{(a) The $\varepsilon$-machine for the process $\ph^+(\past{X},\future{X})$, created by a flipping a biased coin and emitting outcome $2$ when $H \rightarrow T$, $0$ when $T \rightarrow T$, and $1$ when $T / H \rightarrow H$. This process has two causal states $s^+_1$ and $s^+_0$, where the latter includes all pasts ending in either $0$ or $2$. (b) The time-reversed process $P^-(\past{Y}, \future{Y})$. Here pasts ending in $0$, $ 1$ and $2$ now all lead to qualitatively different future behaviour and must be stored in distinct causal states $s^-_0$, $s^-_1$ and $s^-_2$ respectively which occur with respective probabilities $\pi_0^- = (q-pq)/(p+q)$, $\pi_1^- = \pi^+_1 = p/(p+q)$ and $\pi_2^- = pq/(p+q)$. \label{fig:epsilonmachine}}
\end{figure*}

Causal asymmetry captures the discrepancy in statistical complexity when a process is viewed in forward versus reverse time~\footnote{Note that in computational mechanics literature, causal asymmetry is often referred to as causal irreversibility \cite{crutchfield2009time}. Here, we choose the term causal asymmetry to avoid confusion with standard notions of irreversibility used in the physics community}. Consider an observer that encounters $X_{t+1}$ \emph{before} $X_{t}$. Their observations are characterized by the time-reversed stochastic process $\mathcal{P}^- = P^-(\past{Y},\future{Y})$ where past and future are interchanged, such that $\past{Y} = \dots X_{1} X_{0}$, while $\future{Y} = X_{-1}X_{-2} \dots$ and $Y_{t} = X_{-(t+1)}$. A causal model for the time-reversed process then corresponds to a \emph{retrocausal} model for the forward process $P(\past{X},\future{X})$. It generates a statistically accurate retrodiction of the conditional past $P(\past{X}|\future{X} = \future{x})$, using only information contained in the future $\future{x}$. The statistical complexity of this time-reversed process $C^-$ (referred to as the \emph{retrodictive statistical complexity} for $\mathcal{P}$) quantifies the minimal amount of causal information we must assign to model $P(\past{X},\future{X})$ in order of decreasing $t$. \emph{Causal asymmetry} captures the divergence $\Delta C = |C^- - C^+|$. When $\Delta C > 0$, a particular temporal direction is privileged, such that modelling the process in the other temporal direction incurs a memory overhead of $\Delta C$.

Note that the definitions above are entropic measures, and thus take operational meaning at the i.i.d. limit -- i.e. modelling $N$ instances of a stochastic process with statistical complexity $C^+$ requires $N C^+$ bits of past  information, in the limit of large $N$. While this is the most commonly adopted measure in computational mechanics, single shot variants do exist. The \emph{topological state complexity} $D^+$, is particularly noteworthy ~\cite{crutchfield1989inferring}. It captures the minimum number of dimensions (max entropy) $\Xi$ must have to generate future statistics. A single-shot variant of causal asymmetry can thus be defined by the difference $\Delta D = |D^- - D^+|$, between the topological state complexities of $\mathcal{P}^+$ and  $\mathcal{P}^-$.  Here, we focus on statistical complexity for clarity. However many of our results also hold in this single-shot regime. We return to this when relevant.

\textbf{Classical models} -- Prior studies of causal asymmetry assumed all models were classical. In this context, causal asymmetry can be explicitly demonstrated using $\varepsilon$-machines, the provably optimal classical causal models~\cite{crutchfield1989inferring,shalizi2001computational}. This involves dividing the set of pasts into equivalences classes, such that two pasts, $\past{x}$ and $\past{x}'$ lie in the same class if-and-only-if they have coinciding future behaviour, i.e., $P(\future{X} |\past{X} = \past{x}) = P(\future{X} | \past{X} = \past{x}')$. Instead of recording the entire past, an $\varepsilon$-machine records only which equivalence class $\past{x}$ lies within -- inducing an encoding function $\varepsilon: \past{\mathcal{X}} \rightarrow \mathcal{S}$ from the space of pasts $\past{\mathcal{X}}$ onto the space of equivalence classes $\mathcal{S} = \{s_i\}$, known as
\emph{causal states}. At each time-step, the machine operates  according to a collection of transition probabilities $T^{x}_{ij}$: the probability an $\varepsilon$-machine initially in $s_i$, will transition to $s_j$ while emitting output $x$. The classical statistical complexity thus coincides with the amount of information needed to store the current causal state
\begin{equation}
C^+_{\mu} = - \sum_i \pi_i \log \pi_i,
\end{equation}
where $\pi_i$ is the probability the past lies within $s_i$. $\varepsilon$-machines are also optimal with respect to the max entropy~\cite{konig2009operational}, such that the topological state complexity $D_\mu$ of a process is the logarithm of the number of causal states~\cite{crutchfield1989inferring}. Despite their provable optimality, $\varepsilon$-machines still appear to waste memory. The amount of past information they demand typically exceeds the amount the past contains about the future -- the mutual information $E = I(\past{X},\future{X})$. Observing an $\varepsilon$-machine's entire future is insufficient for deducing its initial state. Some of the information it stores in the present is never reflected in future statistics and is thus effectively erased during operation. In general, this waste differs between prediction and retrodiction, inducing non-zero causal asymmetry.

\textbf{Examples} -- We illustrate this by examples, starting with the \emph{perturbed coin}. Consider a box containing a single biased coin. At each time-step, the box is perturbed, causing the coin to flip with probability $p$ if it is in heads ($0$), and $q$ if it is in tails ($1$). The coin's state is then emitted as output. This describes a stochastic process $\mathcal{P}_0^+$. As only the last output is necessary for generating correct future statistics, $\mathcal{P}_0^+$ has two causal states, corresponding to the states of the coin. The statistical complexity $h(\pi_1^+)$ thus represents the entropy of the biased coin, where $\pi_1^+ = \frac{p}{p+q}$ is the probability the coin is in heads and $h(x) = -x \log x - (1-x)\log (1-x)$ is the binary entropy. Furthermore $\mathcal{P}_0^+$ is clearly symmetric under time reversal (i.e., $\mathcal{P}_0^+ = \mathcal{P}_0^-$), and thus trivially causally symmetric.

Suppose we post-process the output of the perturbed coin, replacing the first $0$ of each consecutive substring of $0$s with a $2$ (For example, $\ldots 1000110100\ldots$ becomes $\ldots1200112120\ldots$). This results in a new stochastic process, $\ph^+(\Lx,\Rx)$, called the \emph{heralding coin} $\Ph^+$, which also has two causal states, $s^+_1 = \{ \past{x} | x_{-1} = 1\}$ and $s^+_0 =\{\past{x} | x_{-1} \neq 1 \}$. In fact, one can model $\ph^+(\Lx,\Rx)$ by perturbing the same biased coin in a box, and modifying it to output $2$ -- instead of $0$ -- when it transitions from heads to tails (see Fig. \ref{fig:epsilonmachine}). Thus the heralding coin also has classical statistical complexity $C^+_{\mu} = h (\pi_1^+)$.

Its retrodictive statistical complexity, however, is higher. The time-reversed process $\ph^-(\past{Y}, \future{Y})$ represents an alternative post-processing of the perturbed coin - replacing the \emph{last} $0$ in each consecutive substring of $0$s with a $2$. Now, $0$ can be followed by $0$ or $2$, while $1$ can be followed by anything, and  $2$ can only be followed by  $1$, inducing three causal states $s_j^- = \{\past{y} | y_{-1} = j\}$ (see Fig. \ref{fig:epsilonmachine}). This immediately establishes a difference in the number of distinct configurations needed for causal versus retrocausal modelling. Indeed, $\Ph^+$ fields causal asymmetry
\begin{equation}\label{eqn:causal_assymetry}
\Delta C_\mu = C^-_\mu -  C^+_\mu = (1-\pi_1^-) h(\gamma),
\end{equation}where $\gamma = \pi^-_2/(1-\pi^-_1)$ and $\pi^-_j = \ph^-(\past{y} \in s^-_j)$. To understand this asymmetry, note that when modelling $\Ph^+$, we need only know if the previous output was $1$ (i.e., current state of the coin) to decide whether a $0$ should be replaced by a $2$. To model $\Ph^-$ however, one cannot simply look into the `future' to see if the system will output $1$ next. Causal asymmetry thus captures the overhead required to accommodate this restriction.

In general, causal asymmetry can be unbounded. In Appendix \ref{sec:unbounded}, we describe the class of $n$-$m$ \emph{flower processes}, where $C_{\mu}^+$  scales as $O(\log n)$ while $C_{\mu}^-$  scales as $O(\log m)$. $n$ and $m$ can be adjusted independently, allowing construction of processes where $\Delta C_{\mu} > K$ for any given constant $K$. Setting $m = 2$ for example, can yield a process where $C_{\mu}^+$ can be made arbitrarily high, while $C_{\mu}^-\leq \log 3$. When this occurs, the memory overhead incurred for modelling the process in the `less natural' direction scales towards infinity.

\textbf{Quantum Models} -- A quantum causal model is described formally by an ordered tuple $\mathcal Q = (f, \Omega, \mathcal{M})$ where $\Omega$ is a set of quantum states; $f: \past{{\cal X}} \rightarrow \Omega$ defines how each past $\past{x}$, is encoded into a state $f(\past{x}) = \ket{s_{\past{x}}}$ of a physical system $\Xi$; and $\mathcal{M}$ is a quantum measurement process. To model $P(\past{X}, \future{X})$, repeated applications of $\mathcal{M}$ on $\Xi$ must generate correct conditional future behaviour. That is, application of $\mathcal{M}$ on a system $\Xi$ in state $\ket{s_{\past{x}}}$ must (i) generate an output $x$ with probability $P(X_{0} = x |\past{X} = \past{x})$  and (ii) transition $\Xi$ into a new state $f(\past{x}') = \ket{s_{\past{x}'}}$ where $\past{x}' = \past{x}x$, such that  $L$-repeated applications of $\mathcal{M}$ will generate $x_0,\dots, x_{L-1}$ with correct probability $P(X_{0:L}|\past{X} = \past{x})$ for any desired $L \in \mathbb{Z}^+$~\cite{suen2015classical}. The entropy of a model $\mathcal Q$ is given by the von Neumann entropy $S(\rho) = - \mathrm{Tr}(\rho \log \rho)$, where $\rho = \sum P(\past{X} = \past{x}) \ket{s_{\past{x}}}\bra{s_{\past{x}}}$. Thus the quantum statistical complexity $C^+_{q}$ of a process can be computed by minimizing $S(\rho)$ over all valid models \footnote{Note that while these definitions implicitly assume we can only encode onto pure states $\Omega$. We show in the appendix that even when we allow the encoding function to map directly onto mixed quantum states, it does not help -- there is always a pure state quantum causal model with entropy $S(\rho) = C_q^+$ (see Theorem \ref{thrm:pure}).}.

This optimization is highly non-trivial. There exists no systematic techniques for constructing optimal quantum models, or proving the optimality of a given candidate model. To date, $C^+_{q}$, has only been evaluated for the Ising chain~\cite{suen2015classical}. This process, however, is symmetric under time reversal, implying that $\Delta C_\mu$ is trivially zero. Nevertheless recent advances show multiple settings where quantum models outperform optimal classical counterparts~\cite{gu2012quantum,riechers2016minimized,palsson2017experimentally,tan2014towards,monras2016quantum}. In fact, for every stochastic process where the optimal classical models are wasteful (i.e., $C^+_\mu > E$), it is always possible to design a simpler quantum model~\cite{gu2012quantum}. Indeed, sometimes the quantum memory advantage $C^+_\mu - C_q^+$ can be unbounded \cite{garner2017provably}. Could quantum models mitigate the memory overhead induced by causal asymmetry?

\begin{figure}
\includegraphics[width=0.9\columnwidth]{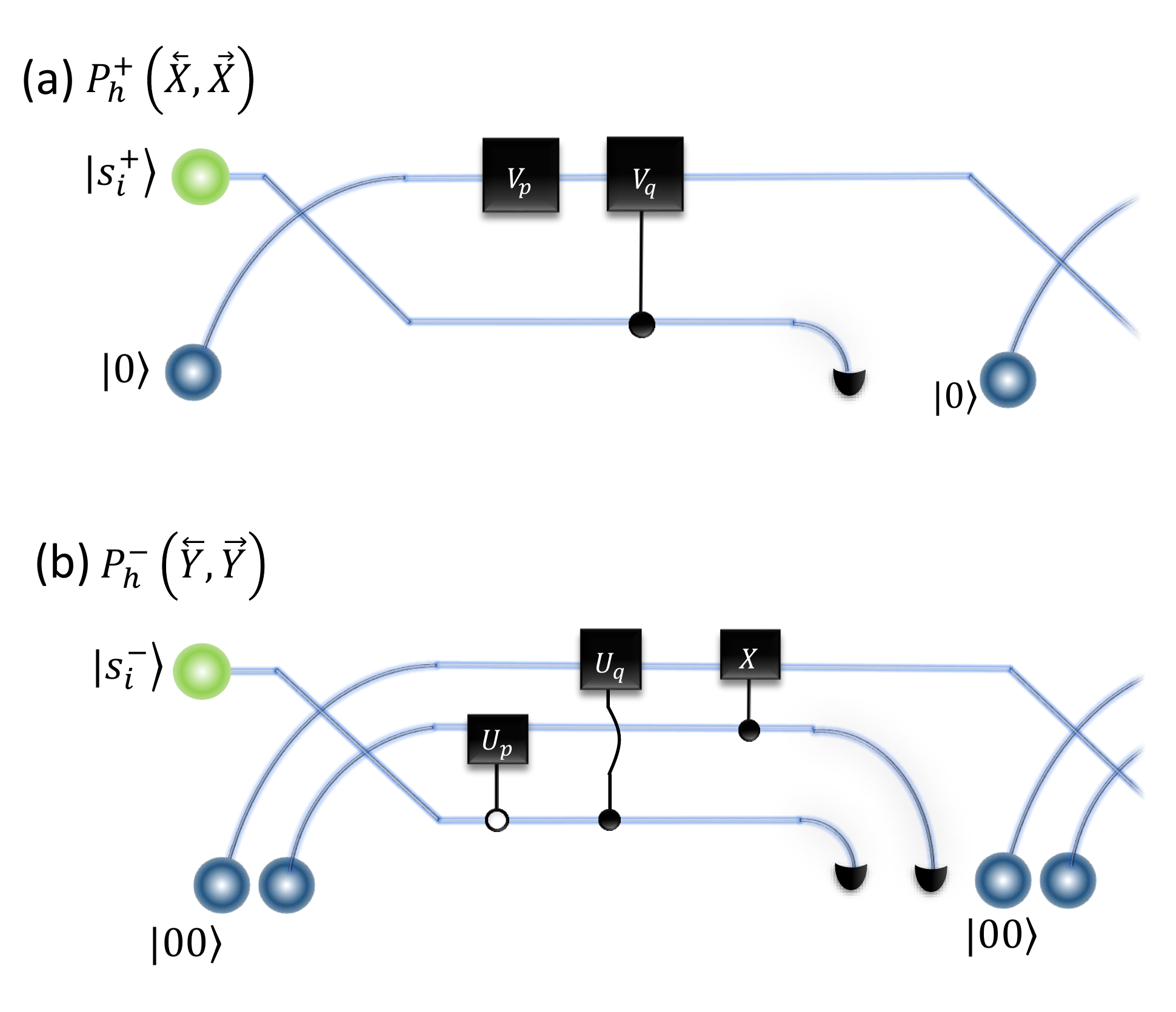}\caption{Quantum circuits for generating (a) $\ph^+(\past{X},\future{X})$ and (b) $\ph^-(\past{Y},\future{Y})$. Here $C_U$ (black circle and line) is the standard control gate  $C_{U}:\ket{w}\ket{\psi} \rightarrow \ket{w} U^{(w \textrm{ mod } 2)} \ket{\psi}$. Meanwhile $\bar{C}_U$ (white circle, black line) is defined as $\bar{C}_{U} \ket{w}\ket{\psi} = \ket{0} U^{(w+1 \textrm{ mod } 2)} \ket{\psi}$. (a) To simulate $\ph^+(\past{X},\future{X})$ we initialize a qubit in state $\ket{s^+_i}$ and an ancilla in state $\ket{0}$. Executing the local unitary $V_p\ket{0} \rightarrow \ket{s^+_0}$, followed by the 2-qubit gate $C_{V_q}$, where $V_qV_p\ket{0} = \ket{s^+_1}$, creates a suitable entangled state -- such that a computation basis measurement of the top qubit yields $x_{t}$, and simultaneously collapses the bottom qubit into the causal state for the next time step. (b) To simulate $\ph^-(\past{Y},\future{Y})$ we prepare state $\ket{s^-_i}\ket{0}\ket{0}$ as input. Execution of $\bar{C}_{U_p}$  where $U_p \ket{0} =  \sqrt{1-p} \ket{0} + \sqrt{p}\ket{1}$, followed by $C_{U_q} $ where $U_q$ satisfies $U_q  \ket{0} = \sqrt{q}\ket{0} + \sqrt{1-q}\ket{1}$, and finally $C_X$ where $X$ is the Pauli X operator generates a suitable entangled state -- such that measuring the first two qubits yields $y_t$ (provided we identify measurement outcome $00 \rightarrow y_{t} = 0$, $10 \rightarrow y_{t} = 1$ and $01 \rightarrow y_{t} = 2$), and collapses the remaining qubit into the quantum causal state for the next time step. In either circuit, retaining only the state of $\Xi$ (green circle) at each time-step is sufficient for generating statistically correct predictions or retrodictions.\label{fig:quantumcircuits}}
\end{figure}

\section{Results}
We study this question via two complementary approaches. The first is a case study of the heralding coin - the aforementioned process that exhibits causal asymmetry. We pioneer methods to establish its provably optimal quantum causal and retrocausal models, and thus produce a precise picture of how quantum mechanics mitigates all present causal asymmetry. The second studies quantum modelling of arbitrary processes with causal asymmetry. Here, $C^+_{q}$ and $C^-_{q}$ cannot be directly evaluated, but can nevertheless be bounded. In doing so, we show that when forced to model such process in the less natural direction, the quantum advantage always exceeds the memory overhead $\Delta C_\mu$.

\textbf{The Heralding Coin} -- Let $\Ph^+$ denote the heralding coin process. Here we first state the optimal quantum models of $\Ph^+$ and $\Ph^-$. We then outline how their optimality is established, leaving details of the formal proof to Appendix \ref{sec:opt_proof}.
The optimal causal model $\mathcal Q^+$ has two internal states;
\begin{eqnarray}\label{eq:forwardsquantumcausalstates}
\ket{s^+_0} &=& \sqrt{1-p}\ket{0} + \sqrt{p}\ket{1}, \notag\\
\ket{s^+_1} & = & \sqrt{q}\ket{2}+ \sqrt{1-q}\ket{1},
\end{eqnarray}
with associated encoding function $\epsilon^+_q(\past{x}) = \ket{s_i^+}$ if-and-only-if $\past{x} \in s_i^+$. Given a qubit in state $\epsilon^+_q(\past{x})$, Fig. \ref{fig:quantumcircuits} establishes the sequential proccedure that replicates expected future behaviour, i.e., samples $\ph^+(\future{X}|\past{X} = \past{x})$.

\begin{figure*}
\includegraphics[width=\textwidth]{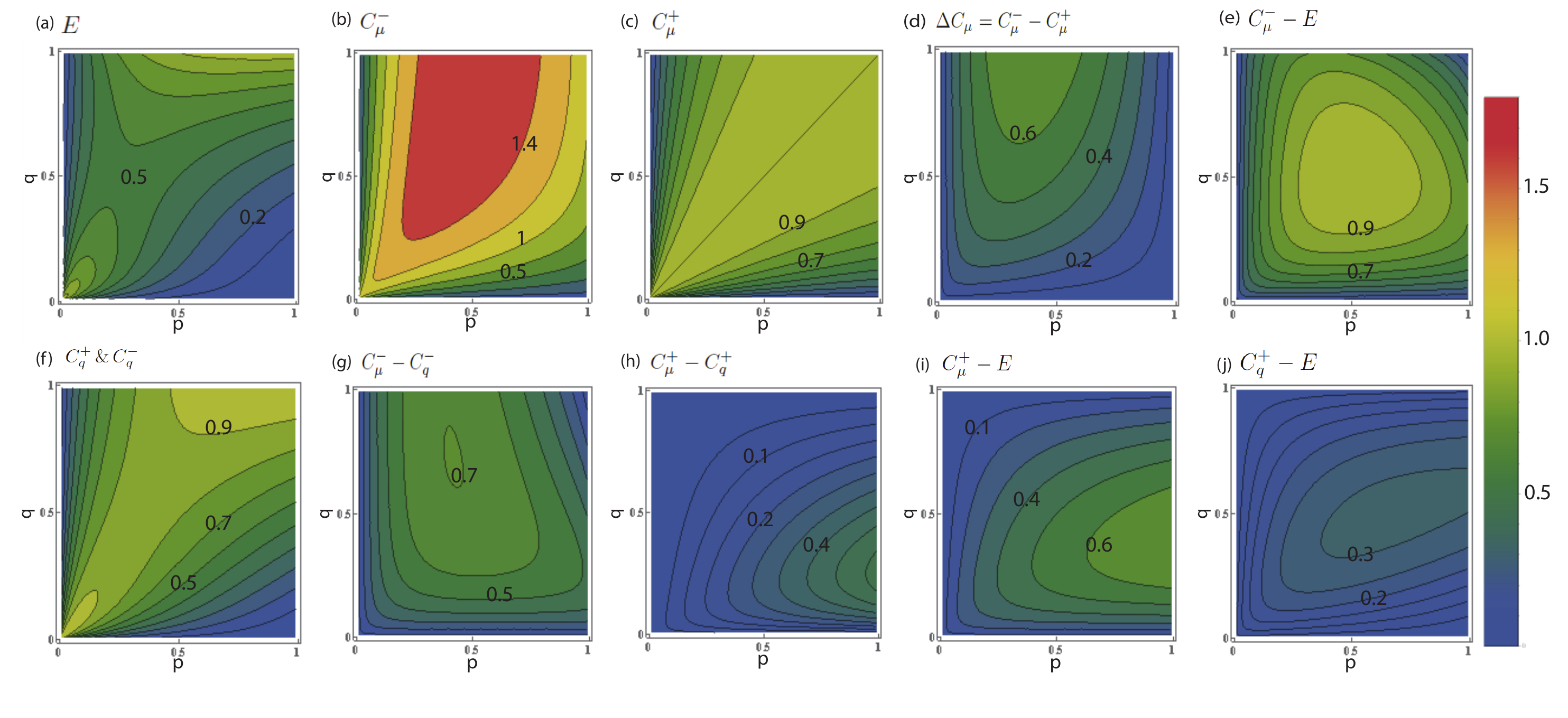}
\caption{Complexity of the heralding coin plotted against $p$ and $q$. The figure illustrates $E \le C_q^{+} = C_q^- \le C_{\mu}^+ \le C_{\mu}^-$ across all values of the parameter space ($0\le p, q \le 1$). (d) depicts the classical causal asymmetry $\Delta C_{\mu} $, and (f) effectively demonstrates $C_q^+ = C_q^-$ and thus $\Delta C_q = 0$. \label{fig:statisticalcomplexity}}
\end{figure*}

Meanwhile the optimal quantum retrocausal model $\mathcal Q^-$ has encoding function $\epsilon_q^-(\past{y}) = \ket{s_i^- }$ if-and-only-if $\past{y} \in s_i^-$, where
\begin{eqnarray}\label{eq:quantumcausalstates}
\ket{s^-_0} &=& \ket{0}, \notag\\
\ket{s^-_1} &=& \sqrt{q}\ket{0} + \sqrt{1-q}\ket{1}, \notag \\
\ket{s^-_2} &=& \ket{1}.
\end{eqnarray}
The associated procedure for sequential generation of $\future{y}$ as governed by $\ph^-(\future{Y}|\past{Y} = \past{y})$ is outlined in Fig. \ref{fig:quantumcircuits}.

To establish optimality, we first invoke the \emph{causal state correspondence}: for any stochastic process with causal states $\{s_i\}$ that occur with probability $\pi_i$, there exists an optimal model $\mathcal Q = (\epsilon_q, \Omega, \mathcal{M})$, where the elements of $\Omega$ are in 1-1 correspondence with $\{s_i\}$ (see Lemma \ref{lem:causalstatecorrespondnace} of Appendix \ref{sec:appendix_definitions}). Since the heralding coin process has two forward causal states, we can restrict
our computation of $C_q^+$ to quantum models where $\Omega = \{\ket{\psi^+_0}, \ket{\psi^+_1}\}$.  Moreover  we can show that the data processing inequality implies $|\langle \psi^+_0 | \psi^+_1\rangle| \leq \sqrt{p(1-q)} \equiv F$ (see Lemma \ref{lem:maxfidelity} of Appendix \ref{sec:appendix_definitions}). The monotonicity between $|\braket{\psi^+_0}{\psi^+_1}|$ and the entropy of the resulting model, together with observation that $|\braket{s^+_0}{s^+_1}| = F$, then implies optimality of $\mathcal Q^+$ (see Theorem \ref{thrm:optimalforward} of Appendix \ref{sec:opt_proof}). This establishes $C_q^+ = S(\rho^+)$ for $\rho^+ = \sum_i \pi_i^+ \ket{s_i^+}\bra{s_i^+}$.

Proving the optimality of $Q^-$ is more involved. First note the causal state correspondence allows us to consider only candidate models $\mathcal Q = (f, \Omega, \mathcal{M})$ where $\Omega = \{\ket{\psi^-_k}\}_{k=0...2}$ has three elements. The data processing inequality can then be used to establish the fidelity constraints $|\langle \psi^-_j |\psi^-_k\rangle| \leq |\langle s^-_j|s^-_k\rangle|$ (see Lemma \ref{lem:maxfidelity} of Appendix \ref{sec:appendix_definitions}). Let $\sigma = \sum \pi^-_k\ket{\psi^-_k}\bra{\psi^-_k}$ with eigenvalues $\lambda_k$, and $\rho^- = \sum \pi^-_k\ket{s^-_k}\bra{s^-_k}$ with eigenvalues $\lambda_k^-$. In Lemma \ref{lem:optimalityinreverse} of Appendix \ref{sec:opt_proof}, we prove that for all choices of $\ket{\psi^-_k}$ satisfying the fidelity constraint $\lambda_k^-$ majorizes $\lambda_k$. Thus $\rho^-$ has minimal entropy among all valid retrocausal quantum models.

$Q^+$ and $Q^-$ exhibit different encoding functions (one maps onto two code words, the other onto three), and invoke seemingly unrelated quantum circuits for generating future statistics (see Fig. \ref{fig:quantumcircuits}). Nevertheless direct computation yields
\begin{equation}
C_q^+ = C_q^- = h\left(\frac{1+\sqrt{c}}{2}\right),
\end{equation}
where $c = (p^2(1+4(1-q)q) -2pq + q^2 )/(p+q)^2$ and $h(\cdot)$ is the binary entropy. Thus $\Delta C_q = 0$ for all values of $p$ and $q$. This establishes our first result:

\begin{result} There exists stochastic processes that are causally asymmetric ($C^+_{\mu} \neq C^-_{\mu}$), but exhibit no such asymmetry when modelled quantum mechanically ($C^+_{q} = C^-_{q}$).
\end{result}

This vanishing of causal asymmetry at the quantum level is not simply the result of saturating the bound given by $E$. Fig. \ref{fig:statisticalcomplexity} shows that $E < C^+_q =C_q^- < C^+_{\mu} < C^-_{\mu}$ for almost all values of $p$ and $q$. While both quantum causal and retrocausal models reduce memory resources beyond classical limits (i.e., $C_q^+  < C^+_{\mu}$ and $C_q^-  < C^-_{\mu}$, see Fig \ref{fig:statisticalcomplexity} f and g), they each still store some unnecessary information ($C^+_q, C_q^- > E$, see Fig.~\ref{fig:statisticalcomplexity}~i).

Our results persist when considering minimal dimensions, rather than minimal entropy required for causal modelling. $\Ph^+$ requires only two causal states, and thus can be modeled using a $2$-level system ($D_\mu^+ = \log 2$). $\Ph^-$, however, has three causal states. Modelling it thus requires a $3$-level system ($D_\mu^- = \log 3$). In contrast, the three quantum causal states of $\Ph^-$ can be embedded within a single qubit, and thus the dynamics of the heralding coin can be modelled using a single qubit in either temporal direction. Therefore this vanishing of causal asymmetry also applies in single shot settings.

\textbf{General Processes} -- We now study quantum mitigation of causal asymmetry for general stochastic processes by bounding $C_q^+$ and $C_q^-$ from above. Let $C_{\mu}^{\min} = \min(C_{\mu}^+, C_{\mu}^-)$ represent the minimum amount of information we need to classically model $P(\past{X},\future{X})$ when allowed to optimize over temporal direction. Meanwhile let $C_{q}^{\max} = \max(C^+_q, C_q^-)$ be the minimal memory a quantum system needs when forced to model the process in the \emph{least favourable} temporal direction. In Appendix \ref{sec:general_process}, we establish the following:

\begin{result} \label{result:upperboundingCq} For any stochastic process $\mathcal{P}$,
\begin{equation}\max(C^+_q, C_q^-) \leq \min(C_{\mu}^+, C_{\mu}^-)
\end{equation}
Equality occurs only if $C_{\mu}^+ = C_{\mu}^- = E$, such that $\mathcal{P}$ is causally symmetric.
\end{result}

Consider any causally asymmetric process $\mathcal{P}$, such that modelling it in the less favourable temporal direction incurs memory overhead $\Delta C_\mu$. Result \ref{result:upperboundingCq} implies that this overhead can be entirely mitigated by quantum models. There exists a quantum model that is not only provably simpler than its optimal classical counterpart, but is also simpler than any classical model of the time-reversed process $\mathcal{P}^-$. In Lemma \ref{thrm:causalsymmetry} (see Appendix \ref{sec:general_process}), we show that such models can be systematically constructed, and align with the simplest currently known quantum models -- $q$-machines~\cite{mahoney2016occam,binder2017practical}. As a corollary, causal asymmetry guarantees both $C^+_q < C_\mu^+$ and $C^-_q < C_\mu^-$, i.e., non-zero quantum advantage exists when modelling in either causal direction.

A variant of these results also applies to topological state complexity. Suppose the number of causal states for $\mathcal{P}$ and its time-reversal $\mathcal{P}^-$ differ, such that $D_\mu^+ \neq D_\mu^-$. Let $D^+_q$ and $D^-_q$ respectively be the logarithm of the minimal dimensions needed to model $\mathcal{P}$ and $\mathcal{P}^-$ quantum mechanically. Appendix \ref{sec:general_process} also establishes that
\begin{result} For any stochastic process $\mathcal{P}$,
\begin{equation}\max(D^+_q, D_q^-) \leq \min(D_{\mu}^+, D_{\mu}^-).
\end{equation}
\end{result}
Given there exists stochastic processes where predictive and retrodictive topological complexity differ (e.g. the heralding coin). This immediately implies the following corollary:
\begin{result} The quantum topological complexity $D_q$ can be strictly less than the classical topological complexity $D_\mu$.
\end{result}
This solves an open question in quantum modelling - whether quantum mechanics allows for models that simulate stochastic processes using not only reduced memory, but also reduced dimensions.

These results have particular impact when $\Delta C_{\mu}$ is exceedingly large. Recall that in the case of the $n$-$2$ flower process, $C_{\mu}^{\min} \leq \log 3$ while $C_{\mu}^+$ scales as $O(\log n)$. Our theorem then implies that $C_{q}^\pm \leq C_{\mu}^{\min} \leq \log 3$. Thus we immediately identify a class of processes whose optimal classical models require a memory that scales as $O(\log n)$, and yet can be modelled quantum mechanically using a single qutrit.

\section{Future Directions}

There are a number of potential relations between causal asymmetry and innovations on the arrow of time, and retrodictive quantum theory. In this section, we survey some of these connections, and highlight promising future research directions.

\textbf{Retrodictive Quantum Mechanics} -- Consider the evolution of an open quantum system that is monitored continuously in time. Standard quantum trajectory theory describes how the system's internal state $\rho(t)$ evolves, encapsulating how our expectations of future measurement outcomes update based on past observations. Retrodictive quantum mechanics introduces the effect matrix $E(t)$ -- a time-reversed analogue of the density matrix $\rho(t)$~\cite{dressel2015weak,gammelmark2013past,wiseman2002weak}. $E(t)$ propagates backwards through time, representing how our expectations of the past change as we scan future measurement outcomes in time-reversed order. The original motivation was that $\rho(t)$ and $E(t)$ combined yield a more accurate estimate of the measurement statistics at time $t$ than $\rho(t)$ alone, allowing improved smoothing procedures~\cite{tan2015prediction,rybarczyk2015forward,weber2014mapping,campagne2014observing}.

While this framework and causal asymmetry differ in motivation and details (e.g. monitoring is done in continuous time, whereas we have so far only considered discrete time), there are also notable coinciding concepts. The standard propagation equation for $\rho(t)$ parallels a causal model for observed measurement statistics, while its time-reversed counterpart governing $E(t)$ parallels a corresponding retrocausal model. It would certainly be interesting to see if such systems exhibit either classical or quantum causal asymmetry. For example, does the resource cost of tracking $E(t)$ differ from that of $\rho(t)$ under some appropriate measure~\footnote{Note that since $\rho(t)$ and $E(t)$ vary over a continuum, the memory costs of tracking either is likely to be unbounded. Thus one may need to modify present approaches. Existing approaches include the use of differential entropies~\cite{marzen2017informational} and studying scaling as we track the process to greater precision~\cite{garner2017provably}}?

Answering these questions will likely involve significant extensions of current results. Our framework presently assumes the process evolves autonomously, and that time is divided into discrete steps. These restrictions will need to be lifted, by combining present results with recent generalizations of classical and quantum computational mechanics to continuum time~\cite{marzen2017informational,elliott2018superior} and input-dependent regimes~\cite{barnett2015computational,thompson2017using,cabello2016thermodynamical}. More generally, such developments will enable a formal study of causal asymmetry in the quantum trajectories formulation of open quantum systems.

\textbf{Arrow of Time in Quantum Measurement} -- Related to such open systems are recent proposals for inferring an arrow of time from continuous measurement~\cite{dressel2017arrow}. These proposals consider continuously monitoring a quantum system initialized in state $\rho_i$, resulting in a measurement record $r(t)$ with some probability $P[r(t)|\rho_i]$. Concurrently, the state of the system evolves through a quantum trajectory $\rho(t)$, into some final configuration $\rho(T) = \rho_f$. The goal is to identify an alternative sequence of measurements, such that for at least one possible outcome record $r'(t)$ occurring with non-zero probability $P[r'(t)|\rho_f]$, the trajectory rewinds. That is, a system initially in state $\rho_f$ will evolve into $\rho_i$, passing through all intermediary states in time-reversed order. An arrow of time emerges as $P[r(t)|\rho_i]$ and $P[r'(t)|\rho_f]$ generally differ, such that one of the two directions occurs with greater probability. An argument via Bayes' theorem then assigns different probabilistic likelihoods towards whether $\rho(t)$ occurred in forward or reverse time.

This framework provides a complementary perspective to our results. It aims to reverse the trajectory of the system's internal state $\rho(t)$, placing no constraints on the relation between the measurement statistics governing $r(t)$ and $r'(t)$. In contrast, causal asymmetry deals with reversing the observed measurement statistics (as described by some stochastic process $\mathcal{P}$), while placing no restrictions on the internal dynamics of the causal and retrocausal models (the two models may even field different Hilbert space dimensions, such as in the heralding coin example).

We also observe some striking parallels. Both works start out with some sequential data, but no knowledge about whether the sequence occurred in forward or reverse time. Both ask the following question: Is there some sort of asymmetry singling out one temporal direction over the other? In the emerging arrow of time from quantum measurement, we are given a trajectory $\rho(t)$, and asymmetry arises from the difficulty (in terms of success probability) of realizing this trajectory in forward versus reverse time. Meanwhile, in causal asymmetry, we are given the observed measurement statistics, and an arrow of time arises from the difference in resource costs needed to realize these statistics causally in forward versus reverse time. It would then be interesting to see if a similar argument via Bayes' theorem can be adapted to causal asymmetry. Supposing more complex machines are less likely to exist in nature (e.g. due to dimensional or entropic constraints), could we then argue whether a given stochastic process is more likely to occur in one causal direction versus the other?

\section{Discussion} Causal asymmetry captures the memory overhead incurred when modelling a stochastic process in one temporal order versus the other. This induces a privileged temporal direction when one seeks the simplest causal explanation. Here we demonstrate a process where this overhead is non-zero when using classical models, and yet vanishes when quantum models are allowed. For arbitrary processes exhibiting causal asymmetry, we prove that quantum models forced to operate in a given temporal order always require less memory than classical counterparts, even when the latter are permitted to operate in either temporal direction. The former result represents a concrete case where causal asymmetry vanishes in the quantum regime. The latter implies that the more causally asymmetric a process, the greater the resource advantage of modelling it quantum mechanically.

Our results also hold when memory is quantified by max entropy. They thus establish that quantum mechanics can reduce the dimensionality needed to simulate a process beyond classical limits. Indeed our results isolate families of processes whose statistical complexity grows without bound, but can nevertheless be modelled exactly by a quantum system of bounded dimension. These features make such processes ideal for demonstrating the practical benefits of quantum models -- allowing us to verify arbitrarily large quantum advantage in single-shot regimes \cite{konig2009operational,dahlsten2013non}, and avoiding the need to measure von Neumann entropy as in current state of the art experiments~\cite{palsson2017experimentally}.

One compelling open question is the potential thermodynamic consequences of causal asymmetry. In computational mechanics, $C_\mu^+$ has thermodynamical relevance in the contexts of prediction and pattern manipulation~\cite{wiesner2012information,garner2017thermodynamics,boyd2017transient,still2012thermodynamics,cabello2016thermodynamical}. For instance, the minimum heat one must dissipate to generate future predictions based on only past observations is given by $W^+_{diss} = k_B T(C_\mu^+ - E)$, where $k_B$ is Boltzmann's constant, $T$ is the environmental temperature, and the excess entropy $E$ is symmetric with respect to time-reversal. Therefore, non-zero causal asymmetry implies that flipping the temporal order in which we ascribie predictions incurs an energetic overhead of $\Delta W_{diss} = k_B T \Delta C_\mu$. In processes where $\Delta C_\mu$ scales without bound, this cost may become prohibitive. Could our observation that $\Delta C_q \le C_\mu^{\mathrm{min}}$  imply such energetic penalties become strongly mitigated when quantum simulators are taken into account?

A second direction is to isolate what properties of quantum processing enable it to mitigate causal asymmetry. In Appendix \ref{sec:general_process}, we establish that all deterministic processes are causally symmetric, such that $C_\mu^{\pm} = C_q^{\pm} = E$ (see Lemma
\ref{lem:determin} of Appendix \ref{sec:general_process}). Randomness is therefore essential for causal asymmetry. Observe also that the provably optimal quantum causal and retrocausal models for the heralding coin both operated unitarily -- such that their dynamics are entirely deterministic (modulo measurement of outputs). Indeed, such unitary quantum models can always be constructed~\cite{binder2017practical}, and we conjecture that this unitarity implies causal symmetry. However, it remains an open question as to whether the optimal quantum model is always unitary.

Insights here will ultimately help answer the big outstanding question of whether the quantum statistical complexity ever displays asymmetry under time-reversal. Identifying any process for which such asymmetry persists implies that Occam's preference for minimal cause can privilege a temporal direction in a fully quantum world. Proof that no such process exists would be equally exciting, indicating that causal asymmetry is a consequence of enforcing all causal explanations to be classical in a fundamentally quantum world.

\textbf{Acknowledgements} -- The authors appreciated the feedback and input received from: Yang Chengran, Suen Whei Yeap, Liu Qing, Alec Boyd, Varun Narasimhachar, Felix Binder, Thomas Elliott, Howard Wiseman, Geoff Pryde, Nora Tischler, Farzad Ghafari and Chiara Marletto. This work was supported by, the National Research Foundation of Singapore and in particular NRF Awards  NRF-NRFF2016-02, NRF-CRP14-2014-02 and RF2017-NRF- ANR004 VanQuTe, the John Templeton Foundation grants 52095 and 54914, Foundational Questions Institute grant FQXi-RFP-1609 and Physics of the Observer grant No. FQXi-RFP-1614, the  Oxford  Martin
School, the Singapore Ministration of Education Tier 1 RG190/17 and the U. S. Army Research Laboratory and the U. S. Army Research Office under contracts No. W911NF-13-1-0390, No. W911NF-13-1-0340, and No. W911NF-18-1-0028. Much of the collaborative was also made possible by the `Interdisciplinary Frontiers of Quantum and Complexity Science' workshop held in January 2017 in Singapore, funded by the John Templeton Foundation, the Centre for Quantum Technologies and the Lee foundation of Singapore.
\begin{appendix}

\section{Technical Definitions} \label{sec:appendix_definitions} We first introduce further technical notation and background that will be used for subsequent proofs.

\begin{definition}[Quantum Causal Model] \label{def1} Consider an ordered tuple ${\mathcal Q} = (f, \Omega, \mathcal{M})$ where $\Omega$ is a set of quantum states; $f: \past{{\cal X}} \rightarrow \Omega$ is an encoding function that maps each $\past{x}$ onto a state $f(\past{x}) = \ket{s_{\past{x}} } $ of a physical system $\Xi$; and $\mathcal{M}$ is a quantum process.  ${\mathcal Q}$ is a quantum model for $P(\past{X}, \future{X})$ if-and-only-if for any $\past{x} \in \past{\mathbfcal{X}}$, whenever $\Xi$ is prepared in $f(\past{x})$  subsequent application of $\mathcal{M}$: (i) generates an output $x$ with probability $P(X_{0} = x |\past{X} = \past{x})$  and (ii) transitions $\Xi$ into a new state $f(\past{x}') = \ket{s_{\past{x}'}}$ where $\past{x}' = \past{x}x$ \cite{suen2015classical}. \end{definition}

Condition (i) guarantees that if a quantum model is initialized in state $f(\past{x})$ then the model's future output $X_0 = x$ will be statistically indistinguishable from the output of the process itself. (ii) ensures the internal memory of the quantum model updates to record the event $X_0 = x$, allowing the model to stay synchronized with the sequence of outputs it has generated thus far. Thus a series of $L$ repeated applications of $\mathcal{M}$ acting on $\Xi$, generates output $x_{0:L} = x_{0} \dots x_{L-1}$ with probability $P(X_{0:L} = x_{0:L} | \past{X} = \past{x})$, and simultaneously transitions $\Xi$ into the state $f(\past{x}x_{0:L} )$. In the limit $L \rightarrow \infty$, the model produces a sequence of outputs $\future{x} = x_{0} x_{1} \dots$ with probability $P(\future{X} | \past{X} = \past{x})$.

The entropy of a quantum model $\mathcal{Q}$ is given by
\begin{equation}
C_q(\mathcal{Q}) = S \left( \rho \right) = - \tr{(\rho \log \rho)},
\end{equation}
where $S(\cdot)$ is the von Neumann entropy, $\rho = \sum_{\past{x}} \pi_{\past{x}} \rho_{\past{x}}$ for  $\rho_{\past{x}} = \ketbra{s_{\past{x}}}{s_{\past{x}}}$, and $\pi_{\past{x}} = P(\past{X} =\past{x})$.
\begin{definition}
$\mathcal{Q}$ is an optimal quantum model for a process $P(\past{X}, \future{X})$, if given any other model $\mathcal{Q}'$, we have  $C_q(\mathcal{Q}')  \ge C_q(\mathcal{Q})$.
\end{definition}

Consider a stationary stochastic process $P(\past{X},\future{X})$, such that $P(X_{0:L}) = P(X_{t:t+L})$ for any $L\in \mathbb{Z}^+$, $t \in \mathbb{Z}$. Let $P(\past{X},\future{X})$ have causal states $\mathcal{S} = \{s_i\}$ each occurring with stationary probability $\pi_i$. Define the conditional distribution $P_i(\future{X}) = P(\future{X}|\past{X} = \past{x} \in s_i)$ as the \emph{future morph} of causal state $s_i$. We will make use of the following two results derived in~\cite{suen2015classical}.

\begin{lemma}[Causal state correspondence]\label{lem:causalstatecorrespondnace} Let $P(\past{X},\future{X})$ be a stochastic process with causal states $\{s_i\}$. There exists an optimal model $\mathcal Q = (\epsilon_q, \Omega, \mathcal{M})$ where $\Omega = \{\ket{s_i}\}$ and $\epsilon_q(\past{x}) = \ket{s_i}$ if-and-only-if $\past{x} \in s_i$.
\end{lemma}

This implies that we can limit our search for optimal models  $\mathcal Q = (f, \Omega, \mathcal{M})$, to those whose internal states $\Omega = \{\ket{\psi_i}\}$ are in one-to-one correspondence with the classical causal states. In addition, it can be shown that $\Omega$ must satisfy the following constraint:

\begin{lemma}[Maximum fidelity constraint] Let $P(\past{X},\future{X})$ be a stochastic process with causal states $\{s_i\}$, and $\mathcal Q = (f,\Omega, \mathcal M) $ be a valid quantum model satisfying  $f(\past{x}) = \ket{\psi_i}$ iff $\past{x} \in s_i$. Then $|\braket{\psi_i}{\psi_j}| \le F_{ij}$, where $F_{ij}=  \sum_{\future{x}}  [P_i(\future{x})P_j(\future{x})]^{\frac{1}{2}}$ is the fidelity between the future morphs of $s_i$ and $s_j$. \label{lem:maxfidelity}
\end{lemma}

These definitions assume that all elements of $\Omega$ are pure. This is because computational mechanics considers only causal models -- models whose internal states do not store more information about the future than what is available from the past. Specifically, let $R$ be a random variable governing the state of a model at $t = 0$.  $I(R,\future{X}|\past{X})$ is then known as the \emph{oracular information}, and represents the amount of extra information $R$ contains about the future $\future{X}$ that is not contained in the past $\past{X}$. For causal models, $I(R,\future{X}|\past{X}) = 0$ \cite{crutchfield2010synchronization}. In Appendix \ref{appendix:mixed}, we show that this allows us to assume all elements of $\Omega$ are pure without loss of generality.

\section{Proofs of Optimality}\label{sec:opt_proof}
Here, we formally prove that the quantum models for the heralding coin given in Eq. (\ref{eq:forwardsquantumcausalstates}) and Eq. (\ref{eq:quantumcausalstates}) are optimal.

\subsection{Optimality of the Causal Model.}\label{appendix:optimal_causal}Let $\Ph^+$ denote the heralding coin process, with corresponding $\varepsilon$-machine depicted in Fig.  \ref{fig:epsilonmachine}(a).
\begin{theorem}\label{thrm:optimalforward}
Consider $\mathcal{Q}^+ = (\varepsilon^+_q, \Omega^+, \mathcal{M}^+)$, where $\varepsilon^+_q (\past{x}) = \ket{s_i^+}$ if-and-only-if $\past{x} \in s^+_i$, with
\begin{eqnarray}\label{eq:forwardsquantumcausalstates2}
\ket{s^+_0} &=& \sqrt{1-p}\ket{0} + \sqrt{p}\ket{1},  \notag\\
\ket{s^+_1} & = & \sqrt{q}\ket{2}+ \sqrt{1-q}\ket{1},
\end{eqnarray}
$\Omega^+ = \{\ket{s_0^+},\ket{s_1^+}\}$, and $\mathcal{M}^+$ described by the quantum circuit in Fig. \ref{fig:quantumcircuits}(a).  $\mathcal{Q}^+$ is an optimal quantum model for $\Ph^+$.
\begin{proof}
We prove this by contradiction. Assume there exists some $\mathcal{Q} = (f, \Omega, \mathcal{M})$ such $C_q(\mathcal{Q}) < C_q(\mathcal{Q}^+)$. Lemma \ref{lem:causalstatecorrespondnace} implies that we can assume $\Omega = \{\ket{\psi_0},\ket{\psi_1}\}$ for some $\ket{\psi_0}$ and $\ket{\psi_1}$ and encoding function $f(\past{x}) = \ket{\psi_i}$ if-and-only-if $\past{x} \in s_i^+$, without loss of generality. $C_q(\mathcal{Q})$, the von-Neumann entropy of the ensemble $\{\ket{\psi_i}, \pi^+_i\}$, is a monotonically decreasing function of $|\langle \psi_0 | \psi_1\rangle|$ \cite{jozsa2000distinguishability}. Thus $C_q(\mathcal{Q}) < C_q(\mathcal{Q}^+)$ implies that  $|\langle \psi_0 | \psi_1\rangle| > |\langle s^+_0 | s^+_1\rangle| = \sqrt{p(1-q)}$. Meanwhile, Lemma \ref{lem:maxfidelity} implies
 \begin{equation}\label{eq:sufficiency}
|\braket{\psi_0}{\psi_1}| \le \sum_{\future{x}} [P^+_0(\future{x})P^+_1(\future{x})]^{\frac{1}{2}} =  \sqrt{p(1-q)}
 \end{equation}
This is a contradiction. Thus no such $\mathcal{Q}$ exists.
\end{proof}
\end{theorem}
\subsection{Optimality of the Retrocausal Model.} \label{appendix:optimal_retrocausal}Let $\Ph^-$ denote the time reversal of the heralding coin process, with corresponding $\varepsilon$-machine in Fig. \ref{fig:epsilonmachine}(b).

\begin{theorem}
Define $\mathcal{Q}^- = (\varepsilon^-_q, \Omega^-, \mathcal{M}^-)$, where $\varepsilon^-_q (\past{y}) = \ket{s_i^-}$ if-and-only-if $\past{y}\in s^-_i$,  with \begin{eqnarray}\label{eq:quantumcausalstates2}
\ket{s^-_0} &=& \ket{0}, \notag\\
\ket{s^-_1} &=& \sqrt{q}\ket{0} + \sqrt{1-q}\ket{1}, \notag \\
\ket{s^-_2} &=& \ket{1},
\end{eqnarray}
$\Omega^- = \{\ket{s_i^-}\}$, and the measurement process $\mathcal{M}^-$ given in Fig. \ref{fig:quantumcircuits}(b). $\mathcal{Q}^-$ is an optimal quantum model for $\Ph^-$.
\end{theorem}

Below we break the proof of this theorem down into a series of small steps. Each step is phrased as a lemma.

\begin{lemma}\label{lem:genericform} Let $\mathcal{Q} = (f, \Omega, \mathcal{M})$ be a quantum model for $\Ph^-$ satisfying $f(\past{y}) = \ket{\psi_i}$ iff $\past{y} \in s_i^-$. Then, up to a unitary rotation,
\begin{eqnarray}\label{eq:parameerization}
\ket{\psi_0} &=& \ket{0}, \notag \\
\ket{\psi_1} &=& r \sin{\theta}\, e^{i\omega} \ket{0} + \sqrt{1-r^2} \,e^{i\alpha}\ket{1} + r\cos{\theta} \ket{2}, \notag \\
\ket{\psi_2} &=& \ket{1},
\end{eqnarray}
for some $\theta \in  [0,\pi/2]$, $0\le r  \le 1$, $\alpha, \omega \in [0,2\pi]$, such that $r \sin \theta \le \sqrt{q}$ and $\sqrt{1-r^2} \le \sqrt{1-q}$.
\begin{proof}Set $F^-_{ij} = \sum_{\future{y}}  [P^-_i(\future{y})P^-_j(\future{y})]^{\frac{1}{2}}$. Explicit evaluation yields $F^-_{01} = \sqrt{q}$, $F^-_{02}  = 0$, $F^-_{12} =  \sqrt{1-q}$. By the maximum fidelity constraint, $|\langle\psi_i | \psi_j\rangle| \le F^-_{ij}$. Thus $\braket{\psi_0}{\psi_2} = 0$. Therefore $\ket{\psi_0} = \ket{0}$ and $\ket{\psi_2} = \ket{1}$ up to a unitary rotation. We can then write $\ket{\psi_1}$ in the form above without loss of generality as the coefficient of $\ket{2}$ can be made real and positive by choosing a suitable definition of basis element $\ket{2}$. Meanwhile constraints on $|\braket{\psi_1}{\psi_2}| \le  \sqrt{1-q}$ and $|\braket{\psi_0}{\psi_1}| \le  \sqrt{q}$ imply $r \sin \theta \le \sqrt{q}$ and $\sqrt{1-r^2} \le \sqrt{1-q}$. \end{proof}
\end{lemma}

Our models described in Eq. \eqref{eq:quantumcausalstates} can be obtained by setting $r \sin \theta \, e^{i\omega} = \sqrt{q}$ and $\sqrt{1-r^2}\,e^{i\alpha} = \sqrt{1-q}$ in Eq. \eqref{eq:parameerization} (i.e.  this corresponds to choosing $\ket{\psi_0} = \ket{s^-_0} $, $\ket{\psi_1} = \ket{s^-_1}$ and $\ket{\psi_2} = \ket{s^-_2}$). The subsequent lemma then establishes that this is the optimal choice.

\begin{lemma}\label{lem:optimalityinreverse} For any quantum model $\mathcal{Q} = (f, \Omega, \mathcal{M})$ of $\Ph^-$ satisfying $f(\past{y}) = \ket{\psi_i}$ if-and-only-if $\past{y} \in s_i^-$.
\begin{equation}
C_{q}(\mathcal{Q}) \ge C_q(\mathcal{Q}^-).
\end{equation}
That is, $\mathcal{Q}^-$, as described by Eq. \eqref{eq:quantumcausalstates}, is the lowest entropy (optimal) model which satisfies the causal state correspondence.
\begin{proof}
By definition $C_q(\mathcal{Q}^-) = S(\rho^-)$ for $\rho^{-} = \sum_i \pi_i^- \ket{s_i^-}\bra{s_i^-}$, where $\pi_i^- = \ph^-( \past{y} \in s_i^-)$ and the states $\ket{s_i^-}$ are given in Eq. \eqref{eq:quantumcausalstates}.  We label the eigenvalues of this state from largest to smallest by $\lambda^{-}_0, \lambda^{-}_1, \lambda^{-}_2$. Meanwhile by the above lemma $C_q(\mathcal{Q}) = S(\rho^{\psi_1})$ where
\begin{eqnarray}\label{eq:mixedstate}
\rho^{\psi_1}&=&  \pi^-_0 \ket{0}\bra{0} + \pi^-_2 \ket{1}\bra{1} + \pi^-_1 \ket{\psi_1}\bra{\psi_1}, \notag
\end{eqnarray}
and $\ket{\psi_1}$ is described by Eq. \eqref{eq:parameerization}. We label the eigenvalues of $\rho^{\psi_1}$ from largest to smallest by  $\lambda^{\psi_1}_0, \lambda^{\psi_1}_1, \lambda^{\psi_1}_2$. To establish that $C_{q}(\mathcal{Q}) \ge C_q(\mathcal{Q}^-)$ it is sufficient to show $\lambda^{-} \succ \lambda^{\psi_1}$, where $\succ$ denotes majorization~\cite{nielsen2001majorization}. This is established by proving that (1) $\lambda^{-}_0 \ge \lambda^{\psi_1}_0$ and (2) $\lambda_0^{-} + \lambda_1^{-} \ge \lambda^{\psi_1}_0 + \lambda^{\psi_1}_1$.

We begin by establishing $\lambda^{-}_0 \ge \lambda^{\psi_1}_0$. By the minimax principle \cite{courant1989methods}, the largest eigenvalue for $\rho^{\psi_1}$ is
\begin{equation}\label{eq:minimax}
\lambda^{\psi_1}_0 = \max_{ |\langle x | x\rangle |^2 = 1} \bra{x} \rho^{\psi_1} \ket{x},
\end{equation}
Suppose that this maximum is attained for some $\ket{x} = \ket{x(t,\phi, \kappa, \eta)}$ such that
\begin{equation}\label{eqn:xmax}
\ket{x}= t \sin{\phi} \, e^{i \eta} \ket{0} + e^{i\kappa} \sqrt{1-t^2} \ket{1}  + t  \cos{\phi} \ket{2},
\end{equation}
where $\phi \in [0,\pi/2]$, $0 \le t \le 1$ and $\eta, \kappa \in [0, 2\pi]$. We can assume the coefficient of $\ket{2}$ is real and positive because Eq. \eqref{eq:minimax} remains unchanged when $\ket{x} \rightarrow e^{i\psi} \ket{x}$. Substituting Eq. \eqref{eqn:xmax} into Eq. \eqref{eq:minimax} yields
 \begin{eqnarray}\label{eq:largesteigenvalue}
 \lambda^{\psi_1}_0 &=& \pi^-_0 |\langle x | 0\rangle |^2 + \pi^-_2|\langle x | 1\rangle|^2  + \pi^-_1 \left| rt \sin{\theta} \sin{\phi} e^{i(\omega - \eta)} \right. \notag \\ && \left. + e^{i(\alpha - \kappa)} \sqrt{1-r^2}\sqrt{1-t^2}    + r t \cos{\theta} \cos{\phi} \right|^2.
  \end{eqnarray}
We defined $\ket{x(t,\phi,\kappa,\eta)}$ to be the vector that maximizes Eq. \eqref{eq:largesteigenvalue}; and thus we have implicitly optimized  over $\kappa$ and $\eta$ in Eq. \eqref{eq:largesteigenvalue}. This optimization  will automatically set  $ e^{i(\alpha - \kappa)}=e^{i(\omega - \eta)} = 1$ (since any two complex numbers $c_1, c_2 \in \mathbb{C}$ satisfy $|c_1 + c_2|^2 \le (|c_1| + |c_2|)^2$). Using this and trigonometry identities to simplify Eq. \eqref{eq:largesteigenvalue}, yields
  \begin{eqnarray}\label{eq:refinedeigenvalue}
\lambda^{\psi_1}_0
&=&  \pi^-_0 t^2 \sin^2 {\phi}+ \pi^-_2 (1-t^2) \notag \\ &&+ \pi^-_1 \left| rt\,\cos{( \phi -\theta)} +  \sqrt{1-r^2}\sqrt{1-t^2}  \right|^2
\end{eqnarray}

We now show that there always exists some $\lambda_0^{\psi_1'}$ such that $\lambda_0^-  \ge \lambda_0^{\psi_1'} \ge \lambda_0^{\psi_1}$. The maximum fidelity constraint implies $r \sin(\theta) \le \sqrt{q}$. Thus there exists some $d\theta$  such that $r \sin (\theta + d \theta) = \sqrt{q}$, (in particular we choose the solution of this equation where $0<\theta + d\theta \le \pi/2$). Consider
\begin{eqnarray}\label{eq:eignevalues}
\lambda^{\psi_1'}_0 &=&  \max_{ |\langle x | x\rangle |^2 = 1} \bra{x} \rho^{\psi_1'} \ket{x}, \notag \\
\rho^{\psi_1'} &=&  \pi^-_0 \ket{0}\bra{0} + \pi^-_2 \ket{1}\bra{1} + \pi^-_1 \ket{\psi_1'}\bra{\psi_1'}, \notag
\end{eqnarray}
where
\begin{eqnarray}
\ket{\psi_1'} &=& r \sin{(\theta + d\theta)} \ket{0} + \sqrt{1-r^2}\ket{1} + r \cos{(\theta + d\theta)} \ket{2} \notag \\
 &=& \sqrt{q}\ket{0} +  \sin{\chi}\sqrt{1-q}\ket{1} +  \cos{\chi}\sqrt{1-q} \ket{2},
 \end{eqnarray}
for $\sin{\chi} =\sqrt{1-r^2}/\sqrt{1-q}$. Furthermore let  $\ket{x'}= \ket{x(t, \beta ,0, 0)}$ be defined as
\begin{equation}
\ket{x'} = t \sin{\beta} \ket{0} + \sqrt{1-t^2} \ket{1} + t\cos{\beta} \ket{2}
\end{equation}
 for $\beta = \min( \pi/2, \phi+d\theta)$.  Then we have
\begin{eqnarray}\label{eq:largesteigenvalueheirachy}
\lambda_0^{\psi_1'}  &\ge& \langle x' | \rho^{\psi_1'} | x'\rangle \notag \\
&=& \pi_0^- t^2\sin^2\beta + \pi^-_2 (1-t^2) \notag \\ &&+ \pi_1^-\left| rt \cos{(\beta -\theta -d\theta )}  + \sqrt{1-r^2}\sqrt{1-t^2}  \right|^2 \notag\\
& \ge &  \pi^-_0 t^2\sin^2{\phi} + \pi^-_2(1-t^2) \notag \\ && + \pi^-_1\left| rt \cos{(\phi -\theta )}  + \sqrt{1-r^2}\sqrt{1-t^2}  \right|^2 \notag \\
& = & \lambda_0^{\psi_1}
\end{eqnarray}
where we have used the fact that $0 \le \phi \le \beta \le \pi/2$  and $|\beta -\theta - d\theta  | \le |\phi -\theta | \le \pi/2$. Specifically these two conditions imply $\sin \beta \ge \sin \phi$ and $\cos{(\beta - \theta -d\theta )} \ge \cos{( \phi -\theta)} \ge 0$. Thus we have $\lambda_0^{\psi_1'} \ge \lambda_0^{\psi_1}$.

To show $ \lambda_0^- \ge \lambda_0^{\psi_1'}$, we define $\ket{y}$ to be the state satisfying $\lambda_0^{\psi_1'} =  \langle y | \rho^{\psi_1'} | y \rangle$. In general we can parameterize
\begin{equation}\nonumber
\ket{y}=we^{i a} \ket{0}+\sqrt{1-w^2}\sin \xi e^{i b}\ket{1} + \sqrt{1-w^2}\cos{\xi}\ket{2}
\end{equation}
for $ 0\le w \le 1$, $\xi \in [0,\pi/2]$, and $a, b \in [0,2\pi]$. Using the same argument as in Eq. \eqref{eq:refinedeigenvalue}, we can show $a = b = 0$ and thus
\begin{eqnarray}
\lambda_0^{\psi_1'}
&=&  \pi^-_0 w^2 + \pi^-_2(1-w^2)\sin^2\xi +  \\ && \pi^-_1 \left| \sqrt{1-q}\sqrt{1-w^2}  \cos{(\chi - \xi)}  +  \sqrt{q}w \right|^2 \notag
\end{eqnarray}

Define $\ket{y'}~=~w~\ket{0}~+~\sqrt{1-w^2}~\ket{1}$. By mirroring the analysis in Eq. \eqref{eq:largesteigenvalueheirachy} we find
\begin{eqnarray}
\lambda_0^{-} &\ge& \bra{y'} \rho^- \ket{y'} \notag \\
&=& \pi_0^- w^2 + \pi_2^- (1-w^2)    \notag \\ && +\pi^-_1   \left| \sqrt{1-q}\sqrt{1-w^2}  +  \sqrt{q}w \right|^2\notag \\
& \ge & \lambda_0^{\psi_1'}
\end{eqnarray}
Together the above results imply $\lambda_0^{-} \ge \lambda_0^{\psi_1'} \ge \lambda_0^{\psi_1}$, establishing step (1) $\lambda_0^- \ge \lambda_0^{\psi_1}$.

For step (2) We must show $\lambda_0^- + \lambda_1^- \ge \lambda_0^{\psi_1} + \lambda_1^{\psi_1}$. However by construction $\rho^-$ only spans a 2-dimensional Hilbert space and thus we have $\lambda_0^- + \lambda_1^- = 1$. It follows that  $\lambda_0^- + \lambda_1^- \ge \lambda_0^{\psi_1} + \lambda_1^{\psi_1}$. Together, these results imply $\lambda^{-}  \succ \lambda^{\psi_1} $ and therefore  $C_q(\mathcal{Q}^-)  \le C_q(\mathcal{Q})$.
\end{proof}
\end{lemma}

By Lemma \ref{lem:causalstatecorrespondnace}, $\Ph^-$ has an optimal quantum model which satisfies the causal state correspondence. Meanwhile by Lemma \ref{lem:optimalityinreverse},  any $\mathcal{Q}$ satisfying the causal state correspondence must have  $C_q(\mathcal{Q}) \ge C_q(\mathcal{Q}^-) $. It follows that $\mathcal{Q}^-$ is an optimal quantum model for $\Ph^-$.

\section{Proof of Result \ref{result:upperboundingCq}}\label{sec:general_process} Here we prove Result \ref{result:upperboundingCq}. To do this, we require some preliminary lemmas. The first connects the capacity for quantum models to improve upon their optimal classical counterparts with causal asymmetry.

\begin{lemma}\label{lemm:connection}If the classical and quantum statistical complexity of a process $\mathcal{P}$ coincide, such that $C^+_q = C^+_\mu$, then $\mathcal{P}$ is causally symmetric and $C^+_\mu = C^-_\mu = E$.
\begin{proof} We first make use of the prior results showing that whenever classical models waste information, more efficient quantum models exist~\cite{gu2012quantum}. Specifically $C^+_\mu > E$ if-and-only-if $C^+_q < C_\mu^+$. Thus, $C^+_q = C^+_\mu$ implies that $C^+_\mu = E$. It is therefore sufficient to show that $C^+_\mu = E$ implies $C^-_\mu = E$.

We prove this by contradiction. Assume $C_{\mu}^+ = E$ but $C_{\mu}^- >E$. Now $C_{\mu}^+ = E$ implies $H(S_{-1} | \future{X}) = 0$, where $S_{-1}$ is the random variable governing the causal state at $t=-1$ \cite{shalizi2001computational}. Thus given $\future{x}$ we can find a unique $s_i$ such that $P( \past{X} = \past{x} | \future{X} =  \future{x})$ is only non-zero when $\past{x}\in s_i$. It follows that the sets $\tau_i = \{ \future{x} | P_i(\future{X} = \future{x}) \neq 0\}$ form a partitioning on the space of all futures (i.e. $\tau_i \cap \tau_j = \varnothing$ for $i \neq j$).

Furthermore any two $\past{x}, \past{x}' \in s_i$ satisfy $P(\future{X} | \past{X} = \past{x}) = P(\future{X} |\past{X}= \past{x}')$, by definition of $s_i$. Thus Bayes' theorem implies that the $\tau_i$ partition the future into equivalence classes $\future{x} \sim \future{x}'$ if-and-only-if $P(\past{X} | \future{X} = \future{x}) = P(\past{X} | \future{X} = \future{x}')$ \footnote{ To derive this simply write $P( \past{X} = \past{x} |\future{X} = \future{x} \in \tau_i) = \frac{ P(\future{X}= \future{x} | \past{X} = \past{x})  P(\past{x}) }{\sum_{\past{x}\in s_i} P (\future{X} = \future{x} | \past{X} = \past{x}) P(\past{x})}= \frac{ P(\future{X}= \future{x}' | \past{X} = \past{x})  P(\past{x}) }{\sum_{\past{x}\in s_i} P (\future{X} = \future{x}' | \past{X} = \past{x}) P(\past{x})}$.}. Hence $\{\tau_i\}$ constitute the retrocausal states. Bayes' theorem also yields $P(\past{X} = \past{x} | \future{X} = \future{x} \in \tau_i) \neq 0$ only when $\past{x} \in s_i$. This implies $H(S_{-1}^- | \past{X} = \past{x}) = 0$, where $S_{-1}^-$ governs the retrocausal state at time $t = -1$. Hence $C_{\mu}^- = E$, which is a contradiction.
\end{proof}
\end{lemma}

%%%%%%%
It follows as a direct corollary of this result that causal asymmetry vanishes for deterministic processes (i.e. processes where $H(\future{X} | \past{X}) =0$).

\begin{lemma} \label{lem:determin}
Any deterministic process $P(\past{X},\future{X})$ has $\Delta C_{\mu} = 0$.
\begin{proof}
Any deterministic process has $E = C_{\mu}^+$ \cite{mahoney2011hidden,shalizi2001computational}. Since $ E \le C_q^+ \le C_{\mu}^+$ it follows that $E = C_{\mu}^+= C_q^+$  and thus according to the above lemma $\Delta C_{\mu} = 0$.
\end{proof}
\end{lemma}
%%%%%%%%

Our next lemma makes use of q-machines \cite{mahoney2016occam}, the simplest currently known quantum models. Consider a process $\mathcal{P} = P(\past{X},\future{X})$ whose classical $\varepsilon$-machine has a collection of causal states $\mathcal{S} = \{s_i\}$ and transition probabilities $T_{ij}^x$. Let $k$ denote the cryptic order of $P(\past{X},\future{X})$, defined as the smallest $l$ such that $H(S_l | X_{0:\infty}) = 0$  \cite{riechers2016minimized,mahoney2016occam,mahoney2011hidden}. The q-machine of $\mathcal{P}$ has internal states $\ket{S_i}$ defined by a recursive relation
\begin{eqnarray}\label{eq:recursivestates}
&& \ket{S_i} = \ket{S_i(l = k)}, \textrm{ where } \\ && \ket{S_i(l)}  = \sum_{xj}\sqrt{T^x_{ij}}\ket{x}\ket{S_j(l -1)},
\end{eqnarray}
and  $\ket{S_i(0)} = \ket{i}$. The associated encoding function satisfies $f(\past{x}) = \ket{S_i}$ whenever $\past{x} \in s_i$ \cite{riechers2016minimized,mahoney2016occam}.
Let $\bar{C}_{q} = S(\rho)$ be the $q$-machine complexity -- the amount of information a $q$-machine stores about the past, where $\rho = \sum_i \pi_i \ket{S_i}\bra{S_i}$. Meanwhile let the max entropy $\bar{D}_{q} = \log \textrm{tr} [\rho]^0$ be the q-machine state complexity --  the minimum dimensionality of any quantum system $\Xi$ capable of storing these internal states.  Note that since q-machines are valid quantum models, $C_q^+ \leq \bar{C}_q^+$ and $C_q^- \le \bar{C}_q^-$. Likewise  $D_q^+ \leq \bar{D}_q^+$ and $D_q^- \le \bar{D}_q^-$. We now establish that the $q$-machine for $\mathcal{P}$ and its time-reversal $\mathcal{P}^-$ have coinciding Von Neumann entropies and coinciding max entropies.

\begin{lemma}\label{thrm:causalsymmetry}Let $P(\past{X},\future{X})$ be a stationary stochastic process and $P^-(\past{Y},\future{Y})$ its time reversal with q-machine complexity $\bar{C}^+_{q}$ and $\bar{C}^-_{q}$, and q-machine state complexity $\bar{D}^+_{q}$ and $\bar{D}^-_{q}$ respectively. Then $\bar{C}^+_{q} = \bar{C}^-_{q}$, and  $\bar{D}^+_{q} = \bar{D}^-_{q}$
\begin{proof}

We first introduce some compact notation. Let $P(\past{X} = \past{x}, \future{X} = \future{x}) = P_{\omni{x}}$ similarly $P(\future{X} = \future{x} |\past{X}=  \past{x}) = P_{\future{x} | \past{x}}$, $P(\future{X} = \future{x} | S_{-1} =  s_i) = P_{i}(\future{x})$ as well as $P(\past{X} = \past{x}) = P_{\past{x}}$ and $P(\past{x} \in s_i) = \pi_i$.

\begin{figure}
\includegraphics[width=0.9\columnwidth]{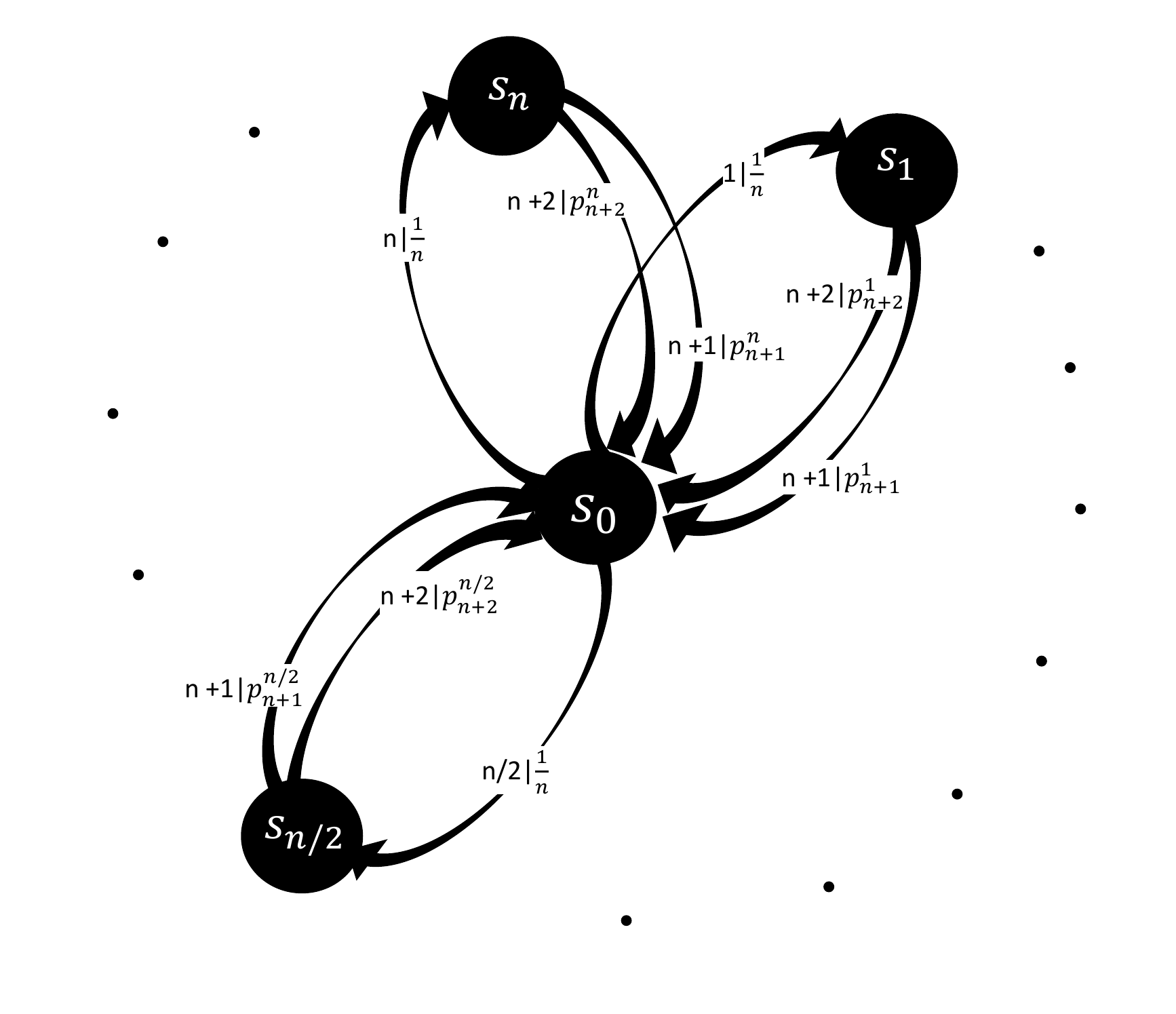}
\caption{The $n$-$m$ flower process, illustrated for the case $m= 2$, and $n$ even.  Physically this process can be generated by a set $\{d_1,\dots, d_n\}$ of $m$-sided dice where each dice $d_i$ is biased, so that it lands on side $j \in \{1,\dots m\}$ with probability $p^{i}_{j}$ (and in general the bias on each dice is different, such that $p^i_j \neq p^k_j$ for $i \neq k$). We randomly select a dice $d_i$, recording the choice $x_t = i$. Afterwards we role the dice, transcribing the outcome $j$ as $x_{t+1} = j+n$. \label{fig:flowerprocess}}
\end{figure}

Now let $\ket{S^+_i}$ denote the internal states of the q-machine for $P(\past{X},\future{X})$, such that $\rho^+ = \sum_i \pi_i \ket{S^+_i}\bra{S^+_i}$, and  $\bar{C}^+_{q} = S(\rho^+)$. From existing work \cite{mahoney2016occam,riechers2016minimized}, we know that $\lim_{l \rightarrow \infty} \langle S_i(l) | S_j(l)\rangle =  \langle S_i(k) | S_j(k)\rangle$. Thus let $\omega^+(l) = \sum_i \pi_i \ket{S^+_i(l)}\bra{S^+_i(l)}$, and $\omega^+ = \lim_{l \rightarrow \infty} \omega^+(l)$ such that $\bar{C}^+_{q} = S(\omega^+)$. Then
\begin{eqnarray}
\omega^+  &=& \lim_{l \rightarrow \infty} \sum_{i}  \pi_i \ket{S_i(l)} \bra{S_i(l)},\notag \\
&=& \sum_{i}  \pi_i  \sum_{\future{x}, \future{x}'}  \sqrt{P_i(\future{x})  P_i( \future{x}')} \ket{\future{x}}\bra{\future{x}'}, \notag \\
&=& \sum_{i} \sum_{\past{x} \in s_i} P_{\past{x}} \sum_{\future{x}, \future{x}'}   \sqrt{P_{ \future{x}|\past{x}}  P_{ \future{x}'| \past{x}}}  \ket{\future{x}}\bra{\future{x}'}, \notag \\
&=& \sum_{\future{x}, \future{x}',\past{x},\past{x}'} \sqrt{P_{ \future{x}|\past{x}} P_{\past{x}} P_{ \future{x}'| \past{x}'} P_{\past{x}'}}  \, \, \delta_{\past{x}, \past{x}'} \ket{\future{x}}\bra{\future{x}'}.
\end{eqnarray}
Furthermore the forward q-machine complexity is given by $\bar{C}^+_{q} = S(\omega^+)$. A similar argument shows that $\bar{C}^-_{q}$ is given by $\bar{C}^-_{q} = S(\omega^-)$, where
\begin{equation}
\omega^- = \sum_{\future{x}, \future{x}',\past{x},\past{x}'} \sqrt{P_{ \past{x}|\future{x}} P_{\future{x}} P_{ \past{x}'| \future{x}'} P_{\future{x}'}}  \, \, \delta_{\future{x}, \future{x}'} \ket{\past{x}}\bra{\past{x}'}.
\end{equation}
Consider now the pure state
\begin{equation}
\ket{\psi}_{\past{X},\future{X}} = \sum_{\past{x}, \future{x}}  \sqrt{P(\past{x}, \future{x})}  \ket{\past{x}, \future{x}}
\end{equation}
which represents that quantum superposition, or q-sample \cite{Aharonov2003}, over all possible output strings of the stochastic process $P(\past{X}, \future{X})$, with associated density operator
\begin{equation}
\rho_{\past{X},\future{X}} =  \sum_{i,i'} \sum_{\past{x} \in s_i} \sum_{\past{x}' \in s_i'}  \sum_{\future{x}, \future{x}'}
\sqrt{P_{\omni{x}'}P_{\omni{x}}}  \ket{\past{x}}\ket{\future{x}}\bra{\past{x}'}\bra{\future{x}'}
\end{equation}
We can verify that $\omega^+ = {\rm Tr}_{\past{X}} [\rho_{\past{X}\future{X}}]$ and $\omega^- = {\rm Tr}_{\future{X}} [\rho_{\past{X}\future{X}}]$. Thus $S(\omega^-) = S(\omega^+)$ and therefore $\bar{C}^+_{q} = \bar{C}^-_{q}$.  The $q$-machine complexity of the forward and backward processes thus coincide.

Note that the rank of $\omega^+$ and $\omega^-$ must also coincide. Thus, an analogous argument establishes that $\log \textrm{tr} [\rho^+]^0 = \log\textrm{tr} [\rho^-]^0$, indicating the two models also have the same dimensionality. Therefore $\bar{D}^+_{q} = \bar{D}^-_{q}$.
\end{proof}
\end{lemma}

We now prove Result \ref{result:upperboundingCq}. Consider any stochastic process $\mathcal{P}$. First assume $\mathcal{P}$ is causally asymmetric, such that $\Delta C_{\mu} \neq 0$. Note first that this implies $C_{\mu}^+, C_{\mu}^- > E$ (by Lemma \ref{lemm:connection}). Meanwhile Lemma \ref{thrm:causalsymmetry} implies that $\bar{C}_q^+ = \bar{C}_q^-$. Thus it is sufficient to show that $\bar{C}_q^+ < C_{\mu}^+$ and $\bar{C}_q^- < C_{\mu}^-$, whenever $C_{\mu}^+, C_{\mu}^- > E$.

Note that for a general process, $\bar{C}_q < C_{\mu}$, if-and-only-if the q-machine has two internal states with non-zero overlap $\langle S_i | S_j \rangle >0$ \cite{nielsen2000quantum}.  It is also previously established that that whenever $C_{\mu} > E$, we can find some $\langle S_i(1)| S_j(1)\rangle > 0$~\cite{gu2012quantum}, as defined by Eq. \eqref{eq:recursivestates}. It follows from the iterative construction that $\langle S_i| S_j\rangle > 0$, and thus $\bar{C}_q < C_{\mu}$. Therefore $C_{\mu}^+ > E $ implies $\bar{C}_q^+ < C_{\mu}^+$ and $C_{\mu}^- > E$ implies $ \bar{C}_q^- < C_{\mu}^-$. Hence, for any causally asymmetric $\mathcal{P}$
\begin{equation}\label{lem:upperboundingCq}
\max(C^+_q, C_q^-) < \min(C_{\mu}^+, C^-_{\mu})
\end{equation}
Conversely, suppose $\max(C^+_q, C_q^-) = \min(C_{\mu}^+, C^-_{\mu})$. Without loss of generality, we can assume $C_{\mu}^+ \leq C_{\mu}^-$. This implies either (i) $C^+_q = C^+_{\mu}$ or (ii) $C^-_q = C^+_{\mu}$. In the case of (i), direct application of Lemma \ref{lemm:connection} implies $C^+_{\mu} = C^-_{\mu} = E$. In the case of (ii) we have $C^+_{\mu} \geq \bar{C}^+_q = \bar{C}^-_q \geq C^-_q = C^+_{\mu}$, which implies $C^+_{\mu} = \bar{C}^+_q$. That is, $q$-machines are not more efficient than $\varepsilon$-machines in modelling $\mathcal{P}$. This is true if-and-only-if $C^+_{q} = C^+_{\mu}$ \cite{gu2012quantum,riechers2016minimized}. Thus Lemma \ref{lemm:connection} again implies $C^+_{\mu} = C^-_{\mu} = E$. This completes the proof.

\section{$n$-$m$ flower process}\label{sec:unbounded}
The family of $n$-$m$ flower processes demonstrate how causal asymmetry can be potential unbounded (see Fig. \ref{fig:flowerprocess}). The process has statistical complexity $C_{\mu}^+ = 1 + \frac{1}{2} \log[n]$. In contrast, the time reversed process will have at most $m+1$ causal states and thus $C_\mu^- \leq \log [m+1]$. Meanwhile the predictive and retrodictive topological state complexities satisfy $D_{\mu}^+ = \log[n+1]$  and  $D_{\mu}^- \leq \log[m+1]$. $n$ and $m$ can be adjusted independently. Setting $m= 2$, and allowing $n \rightarrow \infty$, yields diverging $C_{\mu}^+$ but finite $C_\mu^-$. Thus $\Delta C_{\mu}$ also diverges to infinity. A similar divergence is witness for topological state complexity.

Applying Result 2,  we see that $C^+_q$ and $C^-_q$ are both bounded above by $\log 3$. The same is also true for $D_{q}^+$ and $D_{q}^-$. Thus, quantum models of this process can fit within a single qutrit, whether modelling in forward or reverse time. In the specific case of the former, $C^+_\mu$ and $D^+_\mu$ diverge to infinity. Thus, we obtain a family of processes whose quantum models field unbounded memory advantage - in both the entropic and single-shot sense.

\section{Excluding Mixed State Models}\label{appendix:mixed} In this section we consider more general causal models $\mathcal{Q} = (f,\Omega, \mathcal{M})$ which have the freedom to encode pasts
\begin{equation}
f(\past{x}) = \omega_{\past{x}} = \sum_k q_k(\past{x}) \ket{\psi_k^{\past{x}}} \bra{\psi_k^{\past{x}}},
\end{equation}
into mixed quantum states.  We show that this does not allow for models which are more optimal than those which only encode pasts into pure quantum states.

\begin{theorem}\label{thrm:pure} Consider a stochastic process $P(\past{X},\future{X})$, with a causal model $\mathcal{Q} = (f, \Omega, \mathcal{M})$. If the internal states of $\mathcal{Q}$ are mixed, such that $f(\past{x}) = \sum_i q_i({\past{x}}) \ket{\psi^{\past{x}}_i}\bra{\psi^{\past{x}}_i}$, then we can always find a causal model $\mathcal{Q}' = (f',\Omega', \mathcal{M}')$  such that $f'(\past{x}) = \ketbra{s_{\past{x}}}{s_{\past{x}}}$, and $C_q(\mathcal{Q}') \le C_q(\mathcal{Q})$.
\begin{proof}
Let $P(\past{X},\future{X})$ have causal states $\mathcal{S} = \{s_i\}$. Suppose $\mathcal{Q} = (f, \Omega, \mathcal{M})$ is an optimal causal model for $P(\past{X},\future{X})$, with mixed internal states.

It is trivial to generalize the causal state correspondence to mixed state models. Thus we can assume that $\mathcal{Q}$ has an encoding function where
\begin{equation}
f(\past{x}) = \omega_{i} = \sum_k q_k(s_i) \ket{\psi^{(i)}_k} \bra{\psi^{(i)}_{k}},
\end{equation}
if-and-only-if $\past{x} \in s_i$. So that the internal states $\Omega = \{\omega_i\}$ are in 1-1 correspondence with the classical causal states.

Our proof makes use of the requirement that causal models store no oracular information, i.e. $I(R, \future{X} | \past{X}) = \sum_{\past{x}} P(\past{x}) I(R, \future{X} | \past{X} = \past{x}) = 0$ where $R$ is the random variable governing the memory. Regrouping the pasts into causal state equivalence classes yields   $\sum_{s_i \in \mathcal{S}} \pi_i I(R, \future{X} | \past{x} \in s_i) = 0$, where $\pi_i$ is the probability the past belongs to $s_i$. Thus  $I(R, \future{X} | \past{x} \in s_i) = 0$ for every  $s_i \in \mathcal{S}$.

We have assumed some elements of $\Omega$ are mixed. In particular, suppose we have a specific $\omega_{i'} =\omega \in \Omega$ with $S(\omega) > 0$ that occurs with probability $\pi_{i'} = \pi$. Let $\past{x}$ be a particular past such that $f(\past{x}) = \omega$, and $\Psi = \{\ket{\psi_k}\}$ be a set of pure states that form an unravelling of $\omega$. I.Fe. there must exist some $q_k \in [0,1]$ such that $\omega = \sum_k q_k \ket{\psi_k}\bra{\psi_k}$. Now let $\mathcal{O}_\mathcal{M}$ be a quantum process that maps $\omega$ to a classical random variable $\future{X}$ governed by probability distribution $P(\future{X}|\past{X} = \past{x})$. By definition of a quantum model, this process can always be constructed by concatenations of $\mathcal{M}$ acting on a physical system $\Xi$.

Let $A$ represent the state of $\Xi$ and $B$ be the random variable that governs the resulting output of $\mathcal{O}_\mathcal{M}$ acting on $\Xi$. Zero oracular information implies that $A$ and $B$ must be uncorrelated when conditioned on observing past $\past{x}$. Therefore $\mathcal{O}_\mathcal{M}(\ketbra{\psi_k}{\psi_k}) = \mathcal{O}_\mathcal{M}(\ketbra{\psi_j}{\psi_j}) = \mathcal{O}_\mathcal{M}(\omega) $ for all $\ket{\psi_k}, \ket{\psi_j} \in \Psi$.

Now consider the entropy of $\mathcal{Q}$. By concavity of entropy
\begin{eqnarray}
 S\left( \sum_j \pi_j \omega_j \right)
&=& S\left(\sum_k q_k \left(\pi \ketbra{\psi_k}{\psi_k} + \sum_{\omega_j \neq \omega} \pi_j \omega_j \right)\right) \notag \\
 &\ge&  \sum_k q_k S\left( \pi \ketbra{\psi_k}{\psi_k} + \sum_{\omega_j \neq \omega} \pi_j \omega_j \right) \notag \\
 &\ge&  {\rm min}_{k} \, S\left( \pi \ketbra{\psi_k}{\psi_k} + \sum_{\omega_j \neq \omega} \pi_j \omega_j \right). \notag \\
\end{eqnarray}
 Without loss of generality we can assume this minimum is obtained for $k = 0$. Let $\Omega'' = (\Omega \setminus \omega) \cup \{\ketbra{\psi_0}{\psi_0}\}$ be
 a set of internal states where $\omega$ is replaced with $\ketbra{\psi_0}{\psi_0}$, and define encoding function $f''$ such that $f''(\past{x}) = f(\past{x})$ except when $f(\past{x}) = \omega$, whereby $f''(\past{x}) = \ketbra{\psi_0}{\psi_0}$. Define a new quantum model $\mathcal{Q}'' = (f'', \Omega'', \mathcal{M})$. Clearly $C_q(\mathcal{Q}'') \le C_q(\mathcal{Q})$.

If any of the states in $\Omega''$ are still mixed, then by repeating the above procedure we can replace them with pure states, thereby constructing a model $\mathcal{Q}' = (f',\Omega',\mathcal{M})$ with pure internal states such that  $C_q(\mathcal{Q}') \le C_q(\mathcal{Q})$

\end{proof}
\end{theorem}

\end{appendix}

%\bibliography{time}

\end{document}